\documentclass{emulateapj}

\begin{document}

\title{A Comparison of Ultraluminous X-ray Sources \\
in NGC~1399 and the Antennae Galaxies (NGC 4038/4039)}

\author{Hua Feng and Philip Kaaret}

\affil{Department of Physics and Astronomy, The University of Iowa, Van
Allen Hall, Iowa City, IA 52242; hua-feng@uiowa.edu}

\shortauthors{Feng and Kaaret} \shorttitle{ULXs in NGC 1399 and the
Antennae}

\begin{abstract}

The temporal and spectral properties of ultraluminous X-ray sources
(ULXs, $L_{\rm X}>2\times10^{39}$ ergs s$^{-1}$) and bright X-ray
sources ($L_{\rm X}>3\times10^{38}$ ergs s$^{-1}$) are examined and
compared in two extremely different host environments: the old
elliptical galaxy NGC 1399 and the young, starforming Antennae galaxies
(NGC 4038/4039).  ULXs in NGC 1399 show little variability on either
long or short time scales.  Only 1 of 8 ULXs and 10 of 63 bright
sources in NGC 1399 are variable at a confidence level of 90\%. 
On long timescales, the NGC 1399 sources are steadier than most
Galactic black hole X-ray binaries, but similar to GRS 1915$+$105.  The
outburst duration of the NGC 1399 sources is about 20 yrs, again,
similar to that of GRS 1915$+$105.  The bright X-ray sources in NGC
1399 may be black hole X-ray binaries with giant star companions
similar to GRS 1915$+$105.  The brightest ULX ({PSX-1}) in NGC 1399 is
coincident with a globular cluster, shows a hard spectrum with a photon
index around 1.5, and has a nearly constant luminosity around
$5\times10^{39}$~erg~s$^{-1}$.  It may be an intermediate-mass black
hole (IMBH) in a hard spectral state.  In contrast to NGC 1399, the
ULXs in the Antennae are all variable and a large fraction of the
bright sources (9 of 15) are also variable.  The variability and
luminosity of ULXs in the Antennae suggest they are black hole high
mass X-ray  binaries accreting via Roche-lobe overflow.  A flare with a
duration of about 5~ks is found from Antennae X-42.  The most luminous
ULX, X-16, with a very hard  spectrum ($\Gamma$ = 1.0$\sim$1.3) and
a luminosity which varies by a factor of 10, could be an IMBH
candidate. 

\end{abstract}


\section{Introduction}

Ultraluminous X-ray sources are point-like X-ray sources in external
galaxies with positions off the nucleus and apparent isotropic
luminosities in excess of the Eddington luminosity of a 10$M_\sun$
black hole \citep[$\sim 2\times10^{39}$ ergs s$^{-1}$; for review
see][]{fab06,mil04a}.  ULXs are often found in actively starforming
galaxies, e.g.\ M82 \citep{kaa01} and the Antennae galaxies
\citep{zez02-2}.  There are statistically more ULXs in starburst
galaxies compared to spiral galaxies \citep{kil05}.  \citet{kaa04}
investigated three star burst galaxies and found X-ray sources were
preferentially near star clusters but displaced with higher luminosity
sources nearer to clusters.  All these suggest that ULXs trace young
stellar objects.

Some surveys have reported higher numbers of ULXs in elliptical
galaxies than in spirals \citep{col02}, but the elliptical galaxies
suffer more contamination from background sources \citep{lop06}. 
\citet{irw03} found that bright X-ray sources in elliptical galaxies
have spectra consistent with Galactic X-ray binaries in the high state,
but that the number of sources in the ULX range is consistent with the
number of expected background sources.  However, some elliptical
galaxies contain ULXs associated with globular clusters, e.g.\ NGC 1399
\citep{ang01} and NGC 720 \citep{jel03}, which strongly suggests an
association with the elliptical galaxy.

\begin{deluxetable*}{lllcll}[t]
\tablecaption{Chandra Observations of NGC 1399 and the Antennae. \label{tab:obs}}
\tablehead{
\colhead{Galaxy} & \colhead{index} & \colhead{obs. ID} & 
\colhead{exposure (ks)} & \colhead{Instrument} & \colhead{obs. date}
}
\startdata
 & (a) & 320 & 3.5 & ACIS-I/NONE & 1999 Oct 18 \\
 & (b) & 319 & 55.9 & ACIS-S/NONE & 2000 Jan 18 \\
 & (c) & 239 & 3.6 & ACIS-I/NONE & 2000 Jan 19 \\
NGC 1399 & (d) & 49898 & 13.1 & ACIS-S/HETG & 2000 Jun 15 \\
 & (e) & 240 & 43.5 & ACIS-S/HETG & 2000 Jun 16 \\
 & (f) & 2389 & 14.7 & ACIS-S/HETG & 2001 May 08 \\
 & (g) & 4172 & 44.5 & ACIS-I/NONE & 2003 May 26 \\
\noalign{\smallskip}\hline\noalign{\smallskip}
& (i) & 315 & 72.2 & ACIS-S/NONE & 1999 Dec 01 \\
& (ii) & 3040 & 69.0 & ACIS-S/NONE & 2001 Dec 29 \\
& (iii) & 3043 & 67.1 & ACIS-S/NONE & 2002 Apr 18 \\
The Antennae & (iv) & 3042 & 67.3 & ACIS-S/NONE & 2002 May 31 \\
& (v) & 3044 & 36.5 & ACIS-S/NONE & 2002 Jul 10 \\
& (vi) & 3718 & 34.7 & ACIS-S/NONE & 2002 Jul 13 \\
& (vii) & 3041 & 72.9 & ACIS-S/NONE & 2002 Nov 22 \\
\enddata
\end{deluxetable*}

NGC 1399 and the Antennae are cases, respectively, of an elliptical
galaxy and a starforming galaxy that host a large number of ULXs.
\citet{ang01} compared {\it Chandra} and {\it Hubble Space Telescope}
({\it HST}) images and found 70\% of low mass X-ray binaries (LMXBs) in
NGC 1399 were associated with globular clusters (GCs); the GC
associated sources are brighter than field sources on average.  The
X-ray source population in NGC 1399 may be unusual because a recent
survey of LMXBs in six elliptical galaxies shows that, in general,
there is no significant difference in luminosity between GC associated
and field LMXBs \citep{kim05}.  Several ULXs in NGC 1399 are detected
but their natures are still unclear; they could be IMBHs or multiple
LMXBs \citep{ang01}.  X-ray sources in the Antennae have been well
studied along with a series of observations from {\it Chandra}
\citep{fab01,zez02-2}.  Comparison with optical images shows that the
Antennae X-ray sources are predominately associated with young star
clusters \citep{zez02-3}.  The Antennae ULXs have spectra consistent
with ULXs in nearby galaxies and are not supernova remnants because
they are variable and lack radio emission \citep{zez02-3}. 
\citet{zez02-4} argued that IMBHs could account for a few of the ULXs
in the Antennae but not most of them, which were more likely to be
supercritical accreting stellar mass X-ray binaries due to their
displacement from young star clusters \citep{zez02-2}. The ULX X-37 in
the Antennae has been identified as a background quasar \citep{cla05}.

The most interesting ULXs are potential IMBHs. NGC 1399 and the
Antennae are two extremely different host environments for ULXs. The
comprehensive connection between ULXs and young stellar objects makes
ULXs in NGC 1399 mysterious.  A comparison of ULXs between these two
system may shed light on their natures.  We analyzed {\it Chandra}
archival data for both galaxies (\S~\ref{sec:obs}) in order to measure
their long-term and short-term time variability (\S~\ref{sec:var}) and
spectral variation (\S~\ref{sec:spec}).  We discuss the results and the
implications in \S~\ref{sec:diss}.  We adopt the distance to NGC 1399
as 20.5 Mpc \citep{mer01} and the distance to the Antennae as 13.8 Mpc
\citep{sav04}.  This new distance to the Antennae is obtained by 
measurement of the red giant branch tip and appears accurate to within
2~Mpc, while the older distance was estimated from the recession
velocity.  The new distance is more likely to be accurate.  We note
that this revised distance to the Antennae has reduced the Antennae
source luminosities by a factor of 4 relative to previous X-ray
studies.

\section{Chandra Observations and Data Reduction} \label{sec:obs}

Chandra has made seven observations of NGC 1399 during 1999-2003 and
seven observations of the Antennae during 1999-2002.  A study of some
of the Antennae observations has previously been made \citep{fab03b}. 
The observations are listed in Table~\ref{tab:obs}, in which we
assigned letters to index observations of NGC 1399 and Roman numerals
to observations of the Antennae.  Only data from the Advanced CCD
Imaging Spectrometer (ACIS) were used in this paper.  Observations (a),
(c), (g) were made with the ACIS-I and (b), (d), (e), (f) were from the
ACIS-S but with the High Energy Transmission Grating Spectrometer
(HETGS) applied on (d), (e) and (f).  Observations from (i) to (vii)
were all made with the ACIS-S detector.

Level 2 events files were reduced from level 1 data using CIAO 3.3 with
CALDB 3.2.0.  We created images and exposure maps for each observation
and searched for point-like sources with the {\tt wavdetect} command. 
Exposure maps were produced by assuming a source spectrum of power-law
form with a photon index of 1.5 \citep{ang01,zez02-2} and a Galactic
absorption along the line of sight of $1.34\times10^{20}$ cm$^2$ for
NGC 1399 and $3.83\times10^{20}$ cm$^2$ for the Antennae \citep{dic90}.
A power-law spectrum and a disk blackbody spectrum have similar
luminosities in the 0.3--10 keV band, but will be quite different in
the energy range up to 100 keV. Since our data are only  effective
below 10 kev, an approximation with a single power-law spectrum is
adequate.  In NGC 1399, we selected only sources that were located at
least 20 arcsec off the nucleus and within the D25 circle (6.9 arcmin
in diameter from the RC3 catalog).  In the Antennae, we selected only
sources in the active starforming region, which is bounded by a 1.4
arcmin radius circle.

Since the absolute astrometry accuracy of Chandra is 0.6
arcsec\footnote{radius of the 90\% uncertainty circle, see
\url{http://cxc.harvard.edu/cal/ASPECT/celmon/}}, which is worse than
the angular resolution, a correction for the aspect is necessary
between observations before the same region could be applied in
different observations for source selection.  For NGC 1399, we selected
all detected sources ($>3\sigma$) in the region of interest as defined
above in different observations.  Then we took observation (b) as a
reference and searched in other observations for matched sources that
were located within 1 arcsec of the selected source in (b).  A pointing
offset was calculated by averaging the offsets of all matched sources
for each observation.  A similar correction was applied to observations
of the Antennae with observation (i) used as the reference. 

\subsection{X-ray Source Time Variability} \label{sec:var}

To examine source variability between observations,
luminosity-luminosity (L-L) diagrams are presented for both galaxies.
First, we selected two observations and generated source lists using
{\tt wavdetect}. Then we matched sources and found their luminosities
in both observations. For an unmatched source in the list, we placed
the same elliptical region in the other observation and calculated the
net counts by subtracting the background level, estimated with {\tt
wavdetect}, from the total counts encircled by the ellipse. Then we
derived the luminosity from the net counts normalized with the exposure
map and the galaxy distance. 

\begin{figure*}[!h]
\centering
\plottwo{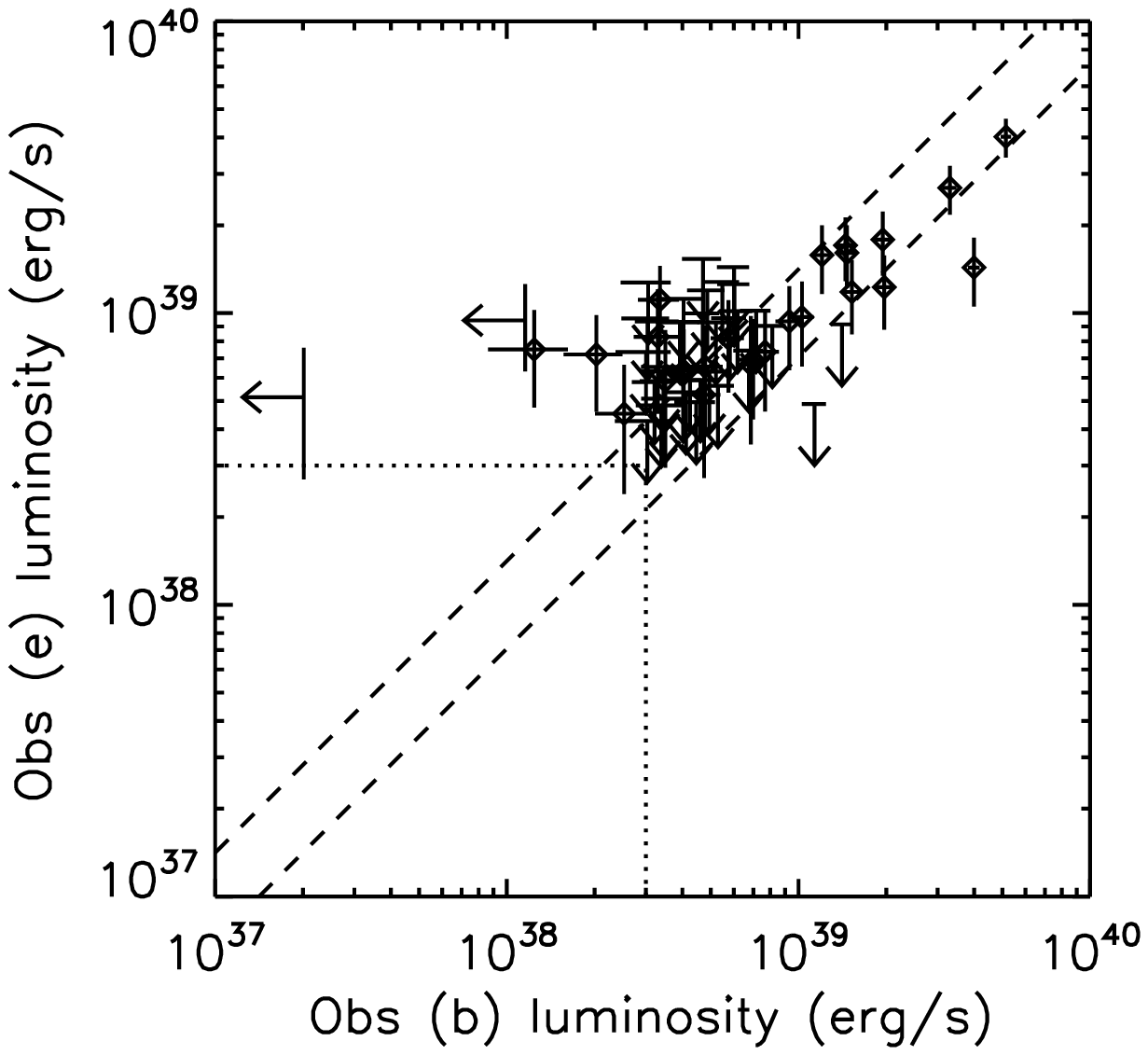}{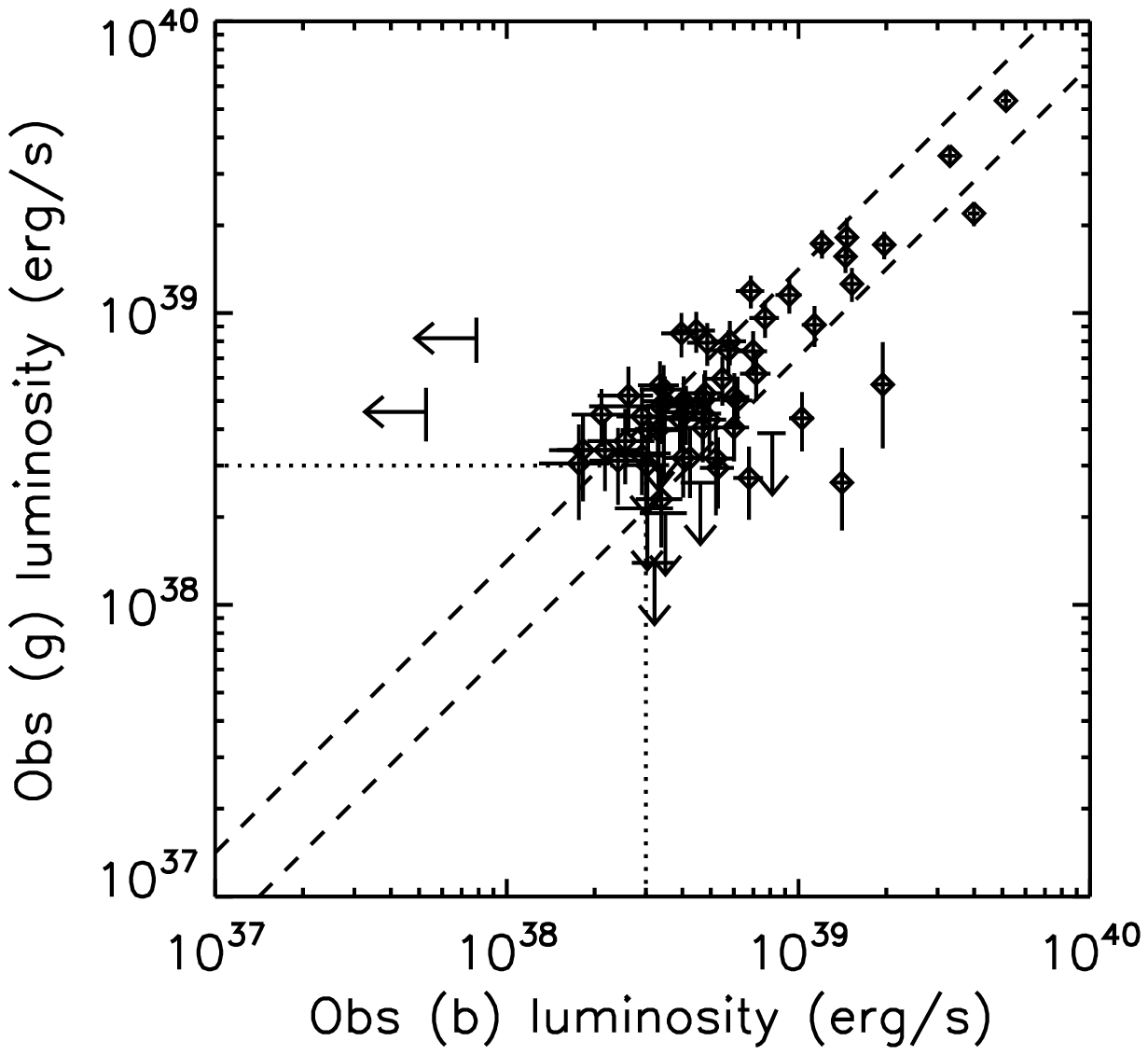}
\caption{The 0.3-10 keV L-L diagrams for X-ray sources above $3\times10^{38}$ ergs
s$^{-1}$ in NGC 1399 between observations (b) and (e) (top), and (b) and (g)
(bottom). The dotted line indicates the lower threshold in luminosity, and the
dashed lines enclose a region without variability considering a systematic error of
$\sqrt{2}$.
\label{fig:lln}}
\end{figure*}

\begin{figure*}
\centering
\includegraphics[width=0.32\textwidth]{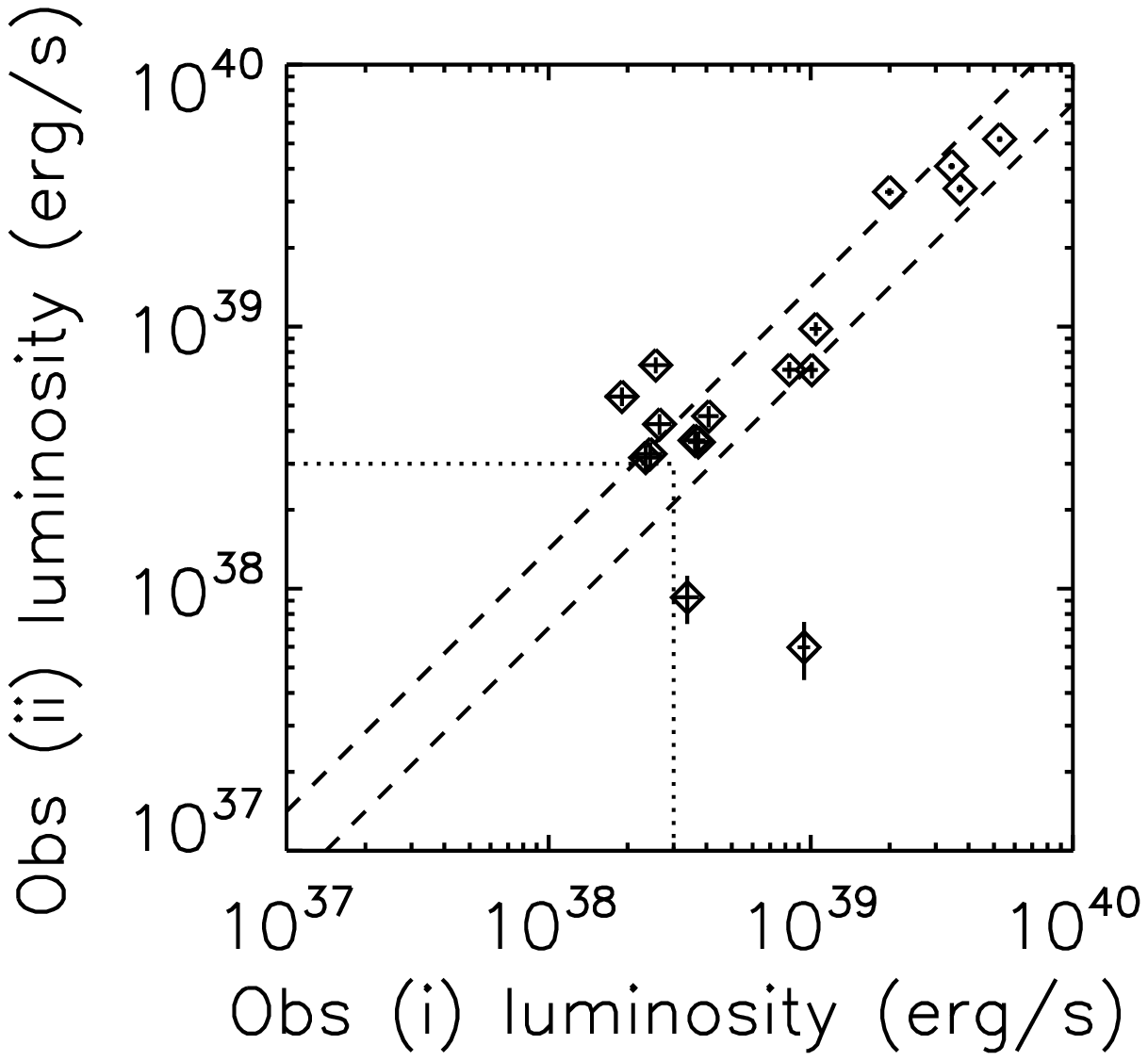}
\includegraphics[width=0.32\textwidth]{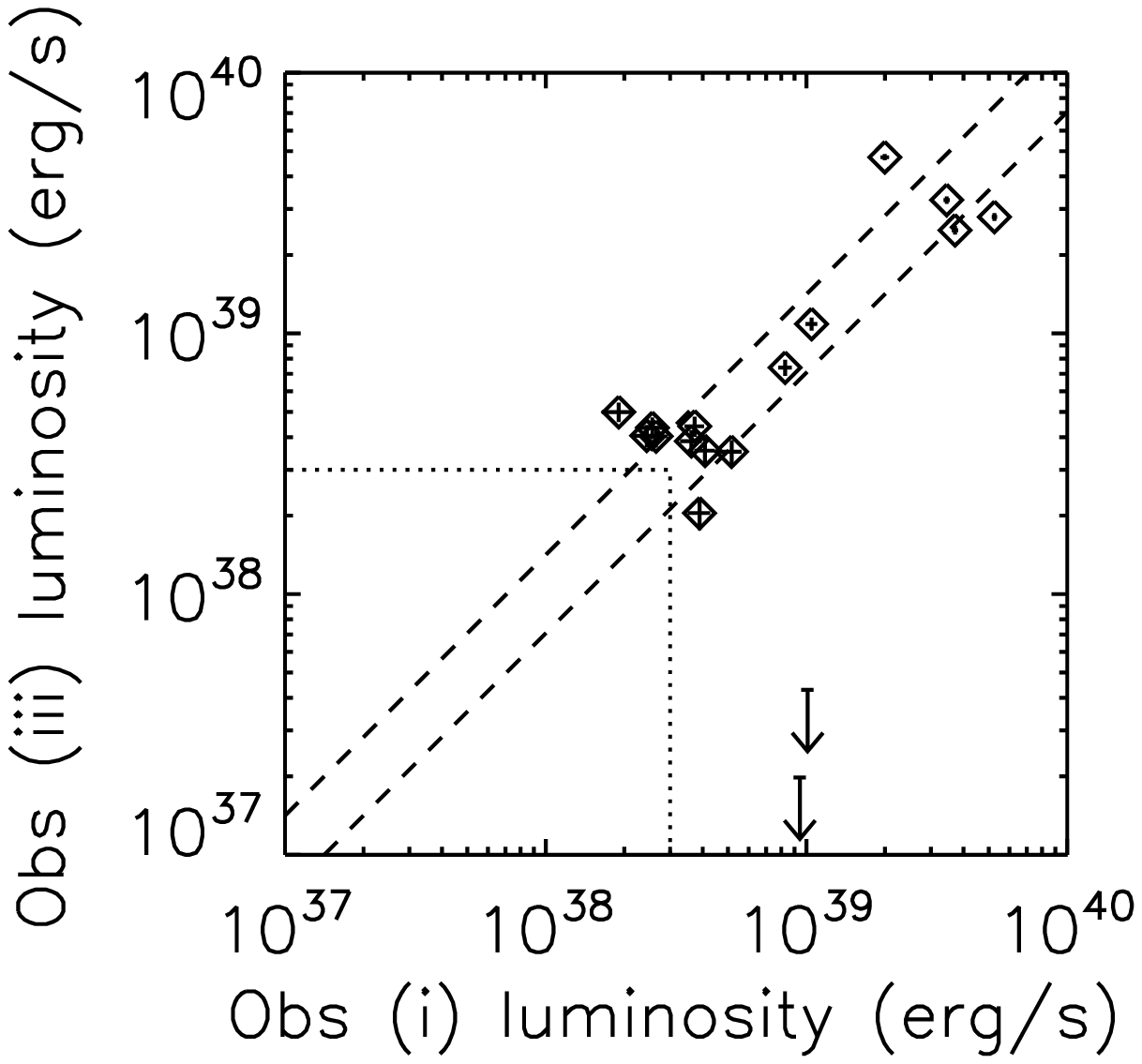}
\includegraphics[width=0.32\textwidth]{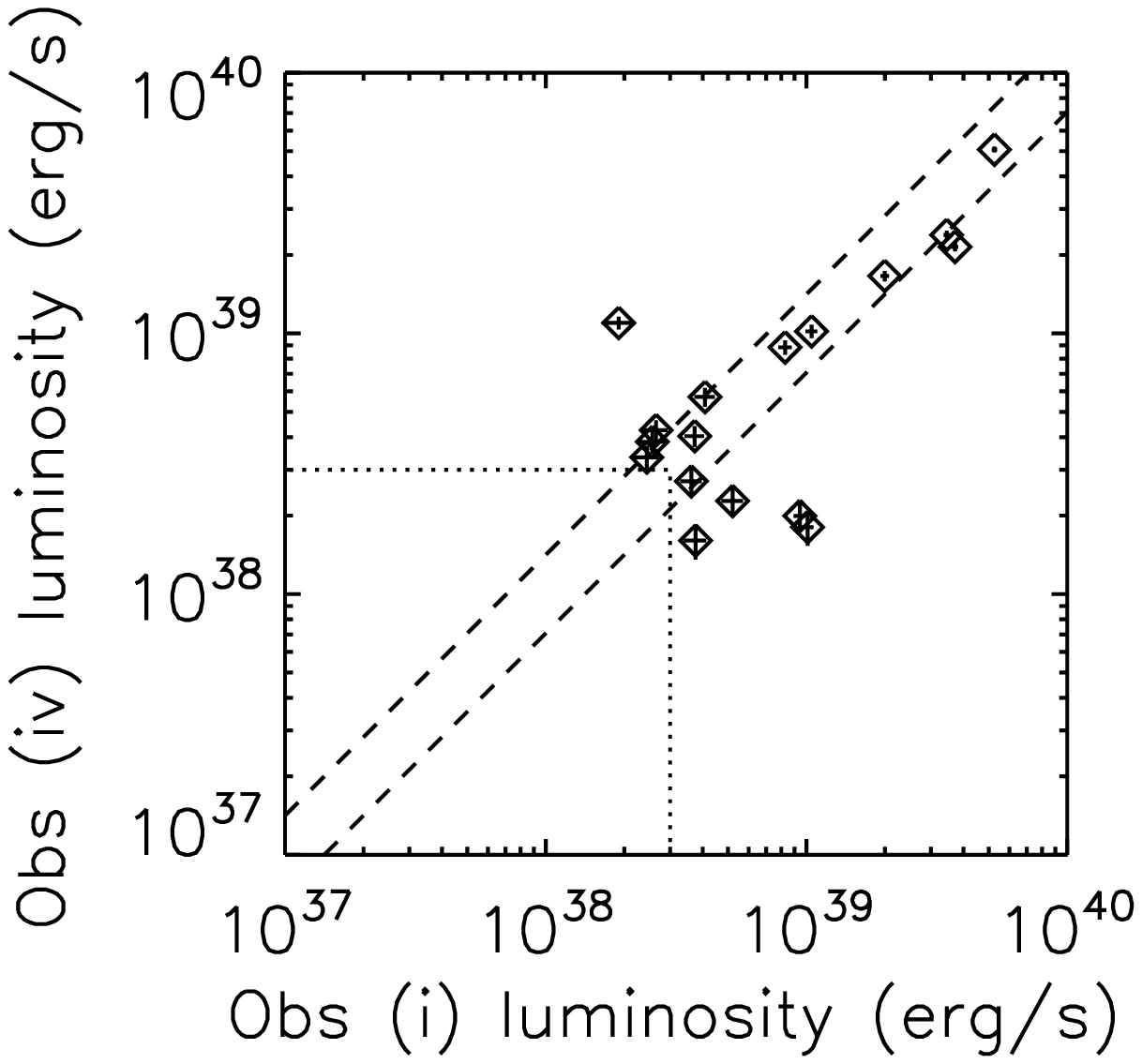}\\
\includegraphics[width=0.32\textwidth]{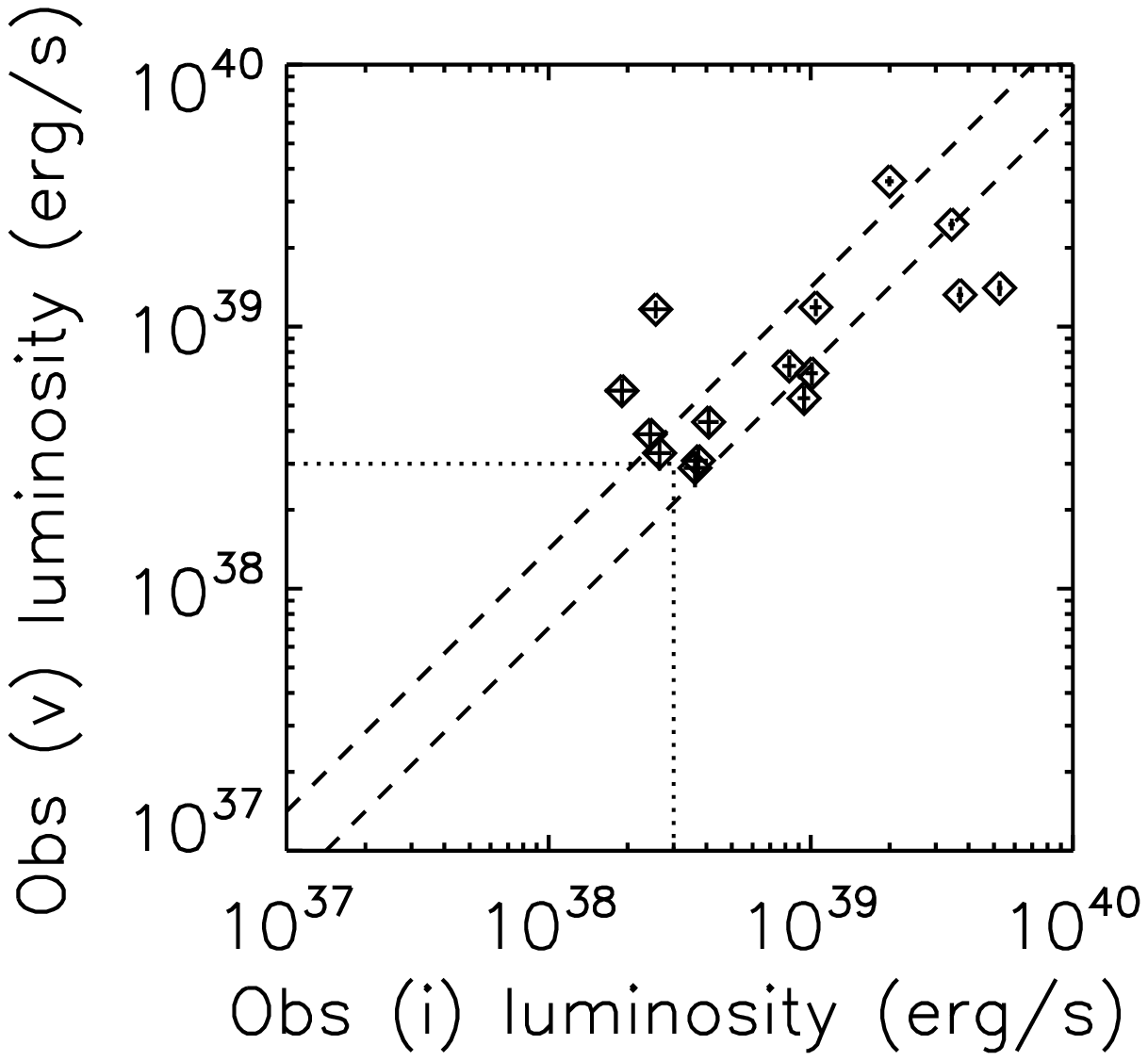}
\includegraphics[width=0.32\textwidth]{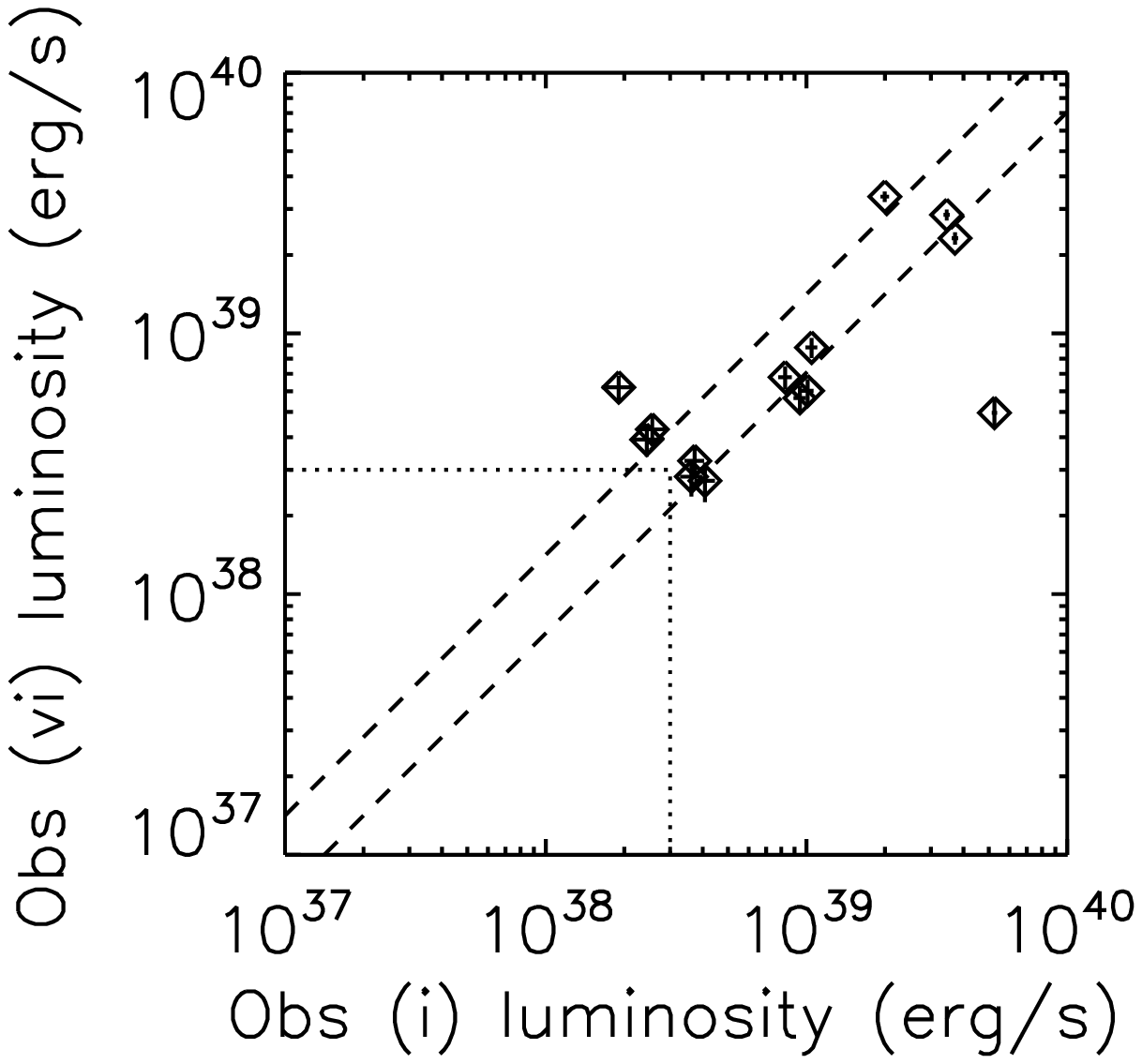}
\includegraphics[width=0.32\textwidth]{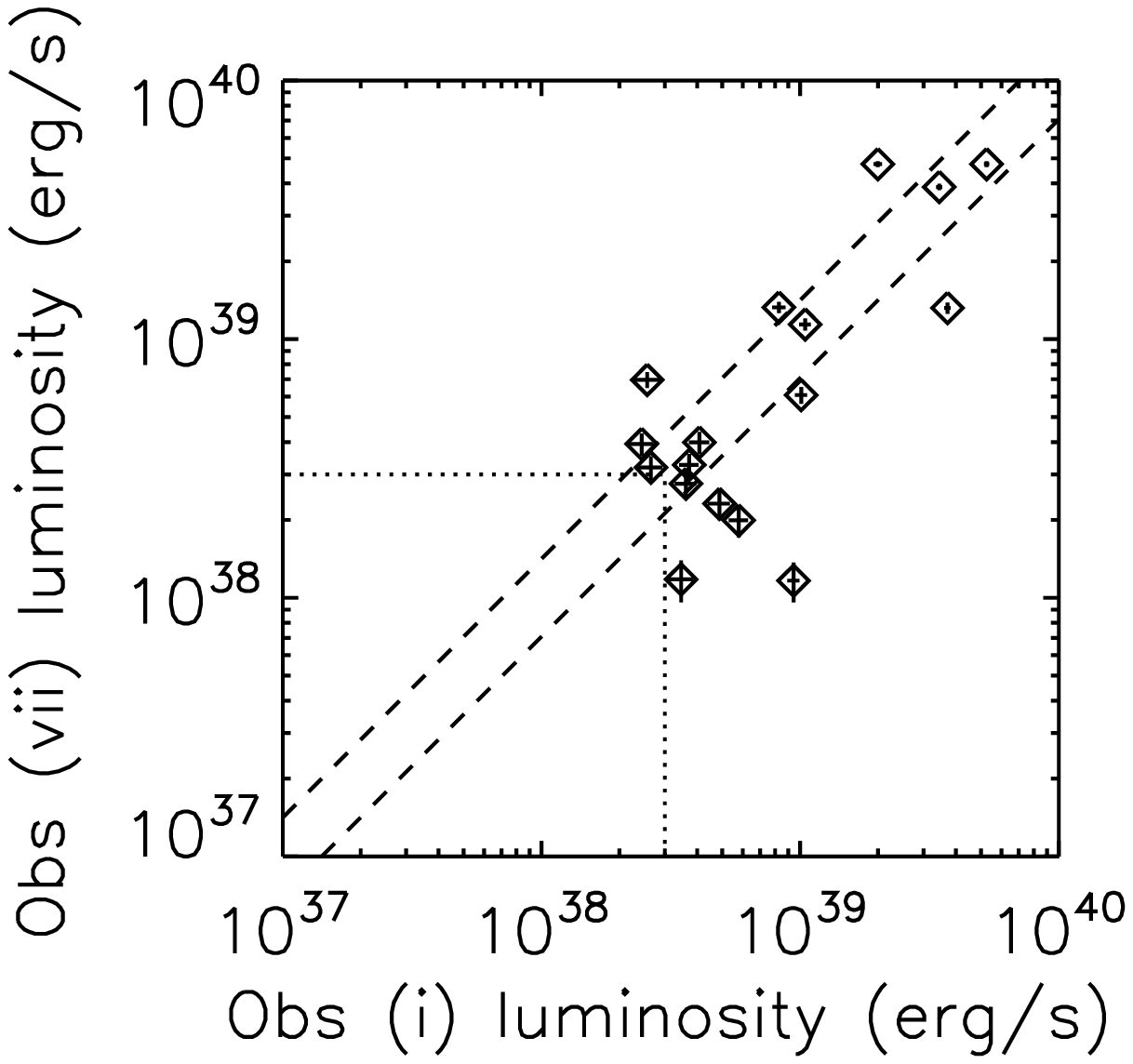}\\
\caption{The 0.3-10 keV L-L diagrams for X-ray sources above $3\times10^{38}$ erg/s
in the Antennae between observation (i) and all others. The dotted line indicates
the lower threshold in luminosity, and the dashed lines enclose a region without
variability considering a systematic error of $\sqrt{2}$. 
\label{fig:lla}}
\end{figure*}

There are three long observations for NGC 1399: (b), (e), and (g). The
L-L diagram between observation (b) and (e) is presented on the top
panel of Fig.~\ref{fig:lln} and that between (b) and (g) is on the
bottom panel of Fig.~\ref{fig:lln}.  A lower threshold of
$3\times10^{38}$ erg~s$^{-1}$ is placed to exclude most neutron stars
as well as faint sources. If the source is undetected with a
significance of $2\sigma$, a $2\sigma$ upper limit is shown. 
Fig.~\ref{fig:lla} shows L-L diagrams for the Antennae sources between
observation (i) and all others.  

\begin{figure*}
\centering
\plottwo{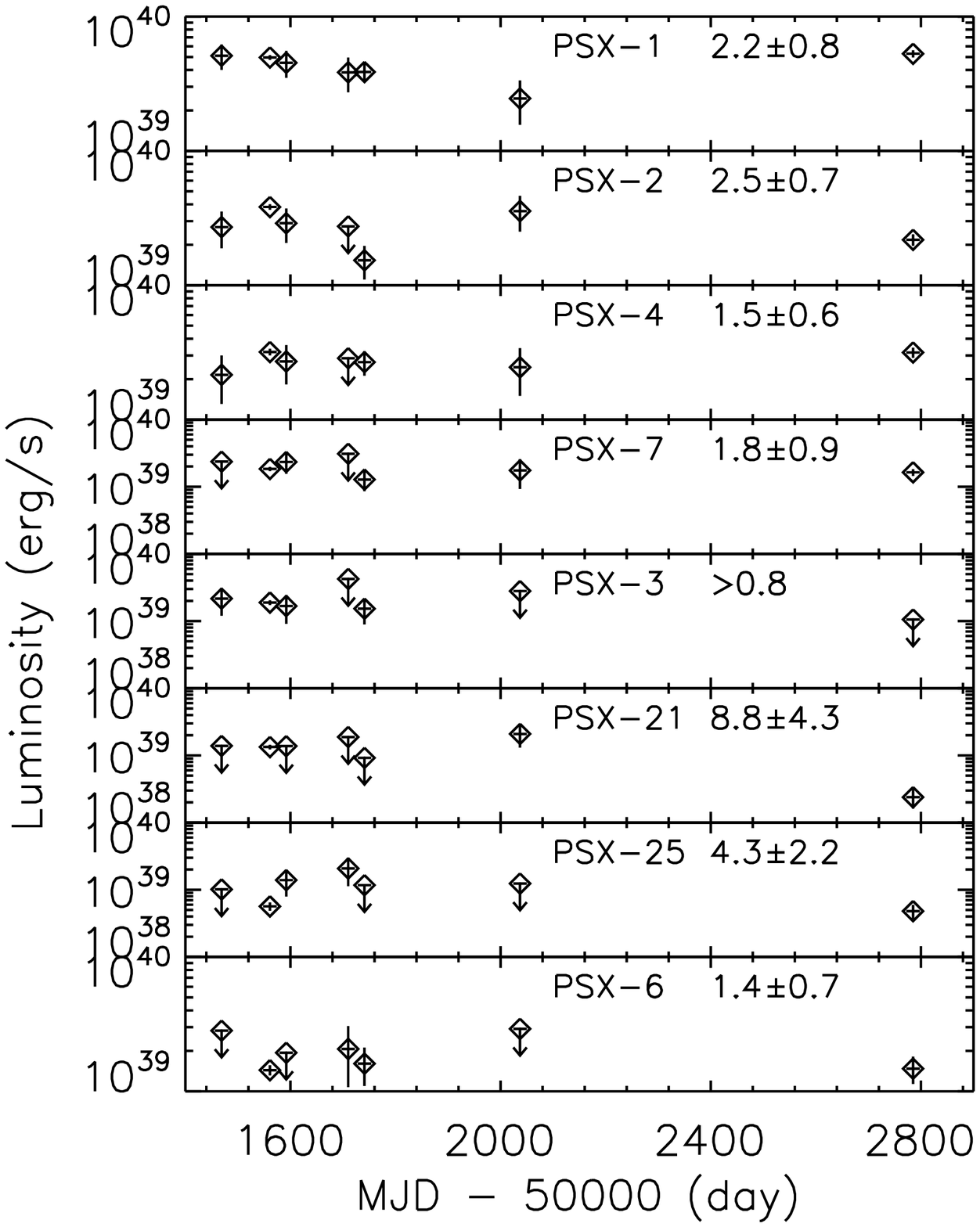}{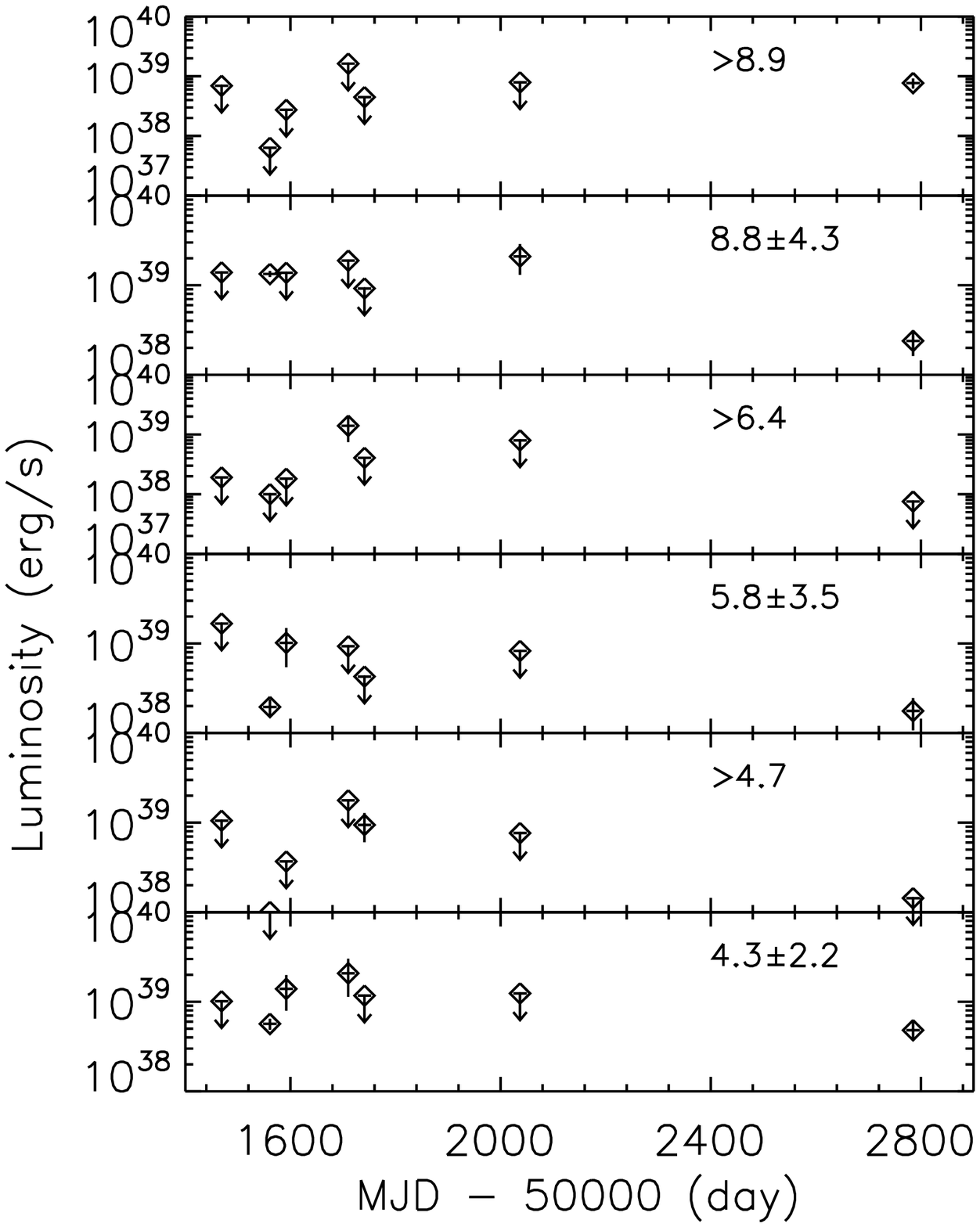}
\caption{Lightcurves of all ULXs (top) and 6 most variable sources (bottom) in NGC
1399. The labels indicate the maximum change in luminosities, which is a number with
a standard error or a 90\% upper limit. The dates for observations (c) and (e) are
shifted to 30 days later for clarity.
\label{fig:lcn}}
\end{figure*}

\begin{figure*}
\centering
\plottwo{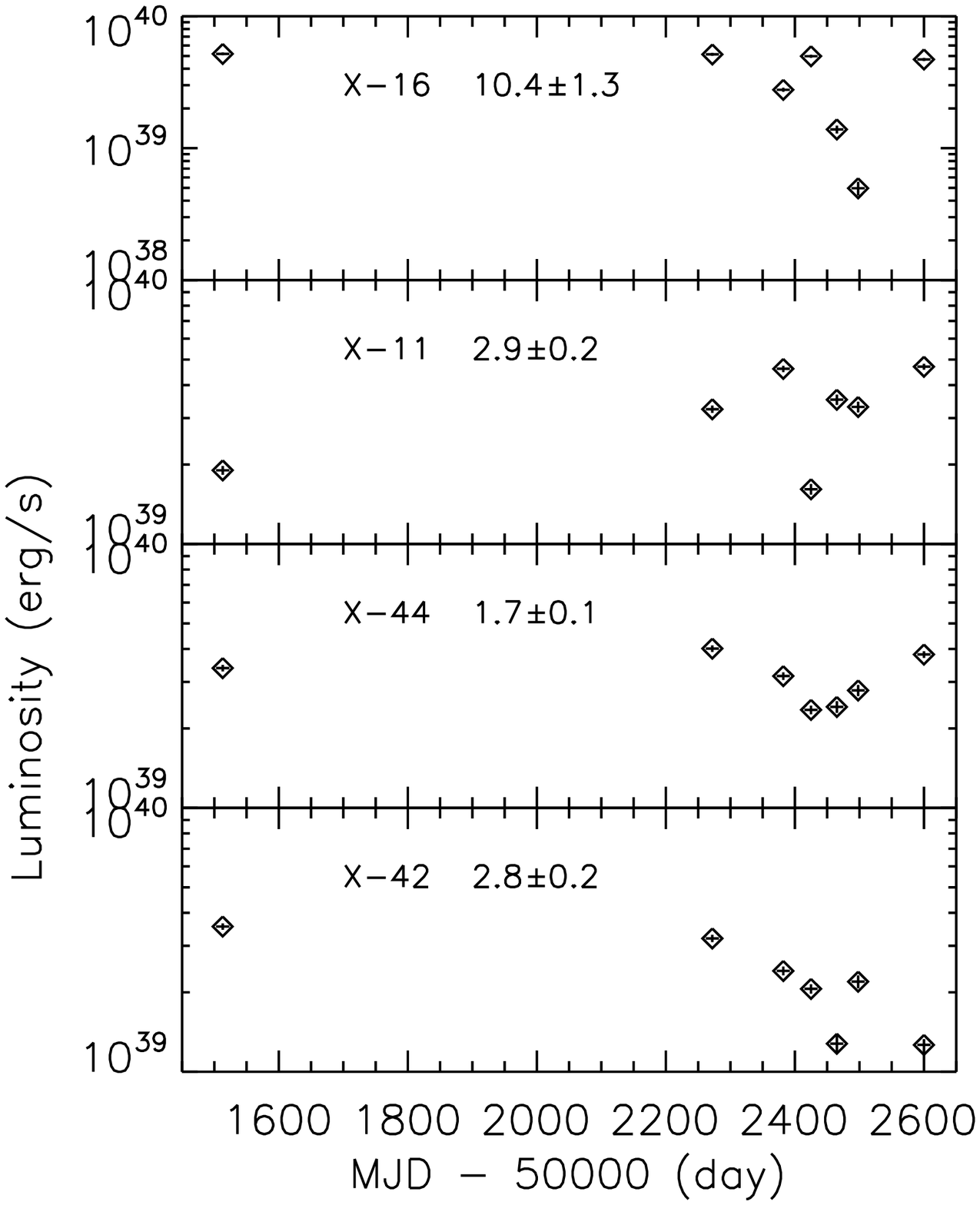}{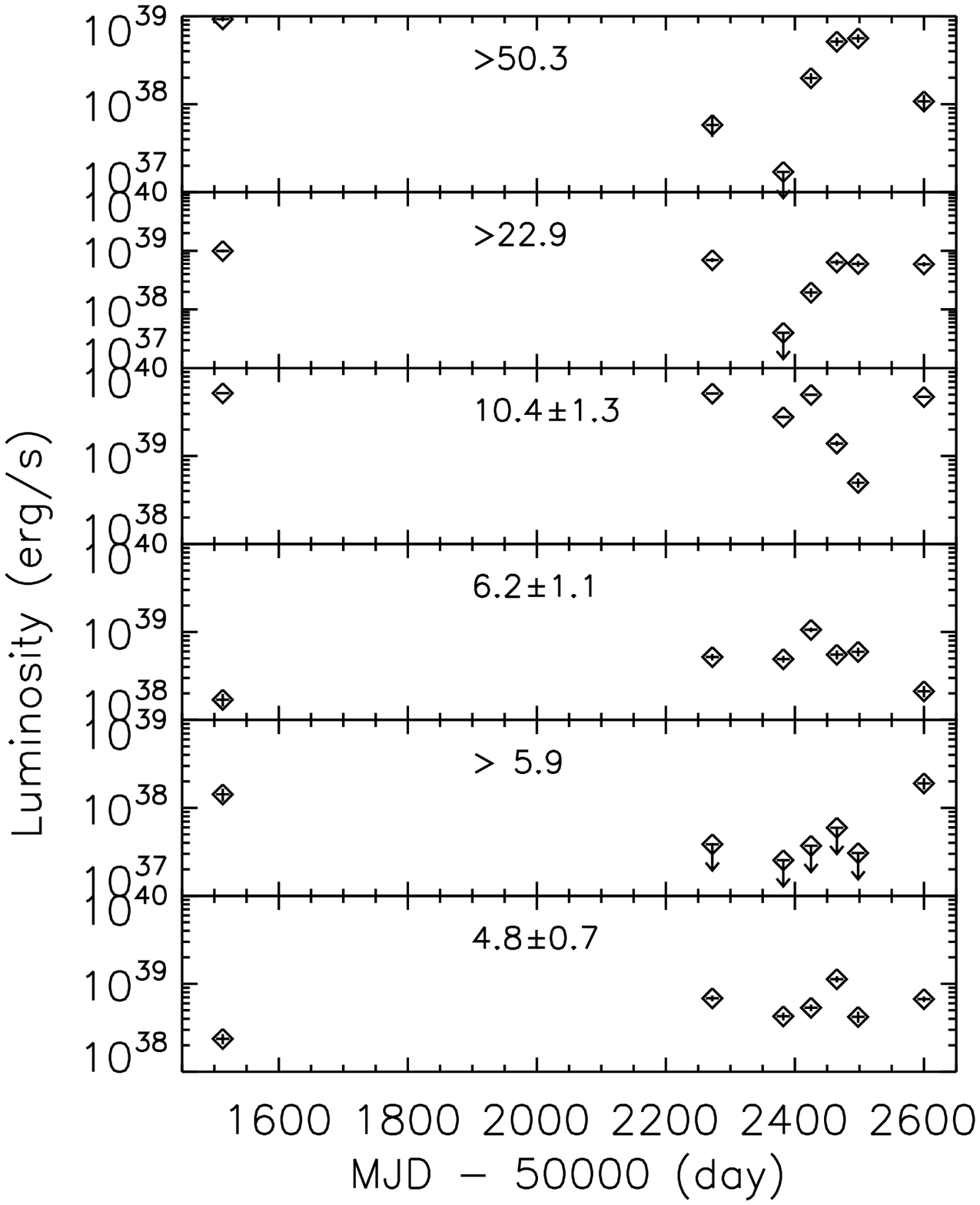}
\caption{Lightcurves of all ULXs (top) and 6 most variable sources (bottom) in the
Antennae. The labels indicate the maximum change in luminosities, which is a number
with a standard error or a 90\% upper limit. The date for observation (vi) is
shifted to 30 days later for clarity.
\label{fig:lca}}
\end{figure*}

We emphasize that we calculate luminosity from counting rate
assuming a fixed spectrum.  Our quoted luminosities are the counting
rate multiplied by a roughly constant factor (the factor is not exactly
constant because different instrument modes were used for NGC 1399 and
the detector response changed slowly for the Antennae).  Because we are
investigating the variability of individual sources, our comparisons of
luminosities is very similar to a comparison of counting rates.  If the
assumed spectra form for a given source is incorrect, then the
luminosities will be incorrect, but all luminosities for any individual
source would be shifted by the same factor (if the spectral shape of an
individual source varies between observations, then that would be
evidence for variability).  We choose to present luminosities instead
of count rates, because the former  are easier to compare with
measurements of other sources.

We note that an estimate of the true luminosity depends on the
spectral form assumed and the energy band used.  The bolometric
luminosity of the sources is likely larger than our quoted values,
since emission is likely present both above and below the 0.3-10~keV
band.  To determine the magnitude of the uncertainty on the luminosity
in our chosen energy band, we fit each source that has enough photons
with an absorbed power-law and measured its intrinsic luminosity (the
photon index and column density are allowed to vary; see details in
Section~\ref{sec:spec}).  Then we fit the spectrum and obtained another
estimate of the luminosity by fixing the absorption to the Galactic
value and the photon index to 1.5.  The ratio between these two
luminosities varies between 0.6$\sim$1.2 for bright X-ray sources in
NGC 1399 and 1$\sim$2 in the Antennae.  Therefore, a bilateral
systematic error of $\sqrt{2}$ is derived.  We also checked with XSPEC
that the systematic error of $\sqrt{2}$ could cover spectral changes
from $\Gamma=1.5-3$ for a single power-law spectrum, $kT=1.5-3$ keV for
a single disk blackbody spectrum, or a reasonable combination of them;
the column density was fixed at the Galactic value for NGC 1399 but was
varied from the Galactic value to 10$^{21}$ cm$^{-2}$ for the
Antennae.  That means this systematic error is valid for black hole
binaries at most spectral states that have been observed in our
Galaxy.  The systematic error is shown as two dashed lines in the L-L
diagram.  Sources lying between the two dashed lines are considered
constant within the accuracy of our measurement.

Lightcurves for the ULXs, which we define as having a luminosity in
0.3-10 keV band above $2\times10^{39}$ ergs s$^{-1}$ in at least one
observation, and for the six most variable sources are shown in
Fig.~\ref{fig:lcn} for NGC 1399 and in Fig.~\ref{fig:lca} for the
Antennae.  The variability is defined as the maximum ratio in
luminosities, which could be a number with a standard error if the
lowest luminosity is detected or a 90\% lower bound if the lowest
luminosity is an upper limit.  The luminosities in the lightcurves are
derived slightly differently than those in the L-L diagram.  First we
generate a complete source list, which is based on detected sources in
the reference observation ((b) or (i)), as well as sources only
detected in other observations.  We place the same circular region for
a given source in all observations and calculate the luminosity.  This
difference in technique causes a mild difference with the luminosities
in the L-L diagrams but is consistent within the errors.  There are 8
ULXs found in NGC 1399, PSX-1, PSX-2, PSX-4, PSX-7, PSX-3, PSX-21,
PSX-25 and PSX-6 \citep[names are from][]{hum03}, sorted by the maximum
luminosity. Among them, PSX-1, PSX-2, PSX-25 and PSX-6 are found to be
associated with GCs.  The GC identifications are from the {\it HST}
image in \citet{ang01} and the source list of \citet{dir03}.  There are
6 ULXs detected in the Antennae, X-16, X-11, X-44, X-42, X-37 and X-29
\citep[names are from][]{zez02-2}, sorted in the same way.  We note
that the Antennae X-37 has been identified as a background quasar
\citep{cla05} and X-29 is the nucleus of NGC 4039 \citep{zez02-3}. We
excluded these two ULXs as well as the nucleus of NGC 4038 (X-25) from
all analysis in our paper.

\begin{figure} 
\centering
\plotone{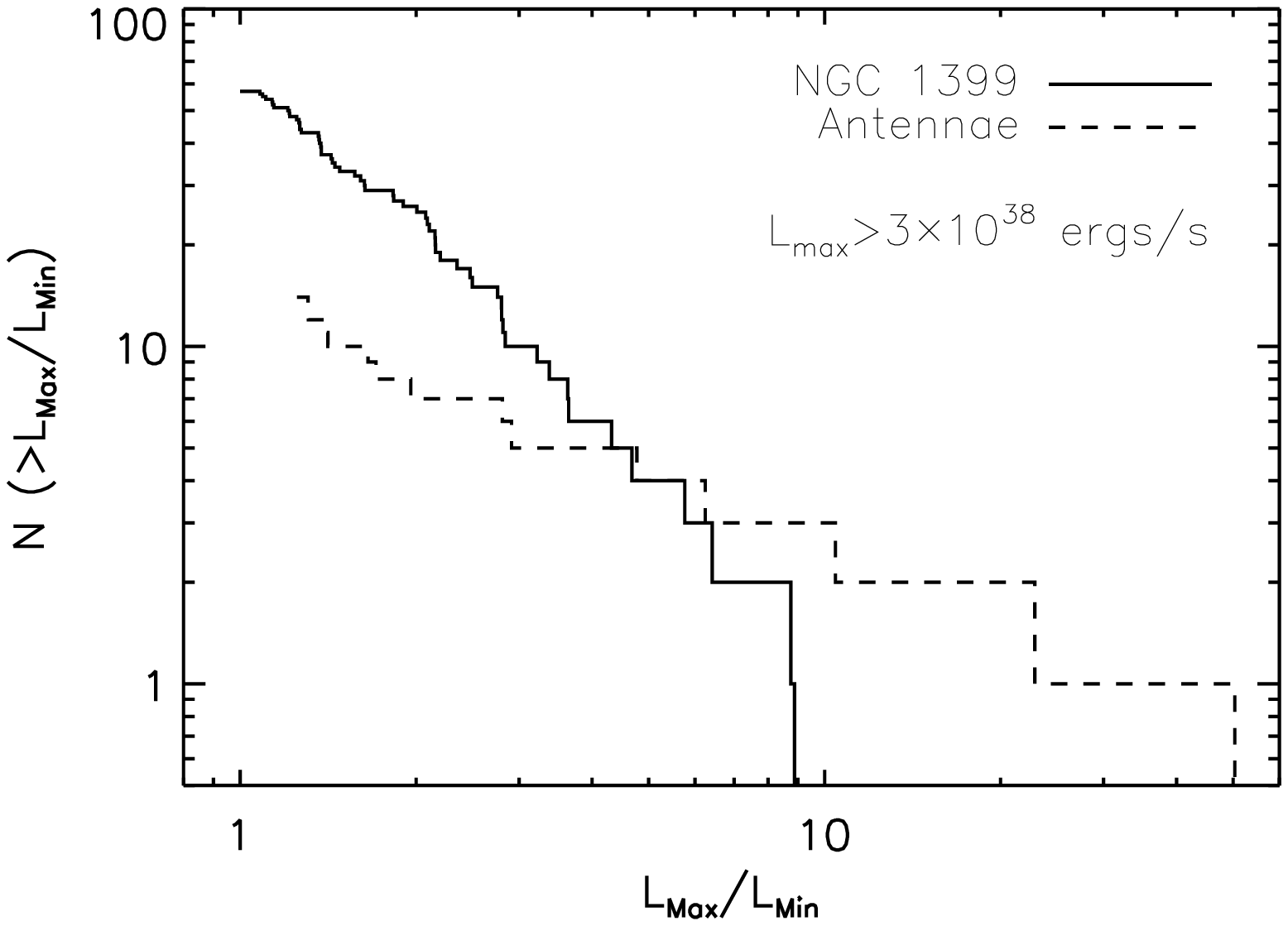}\\
\plotone{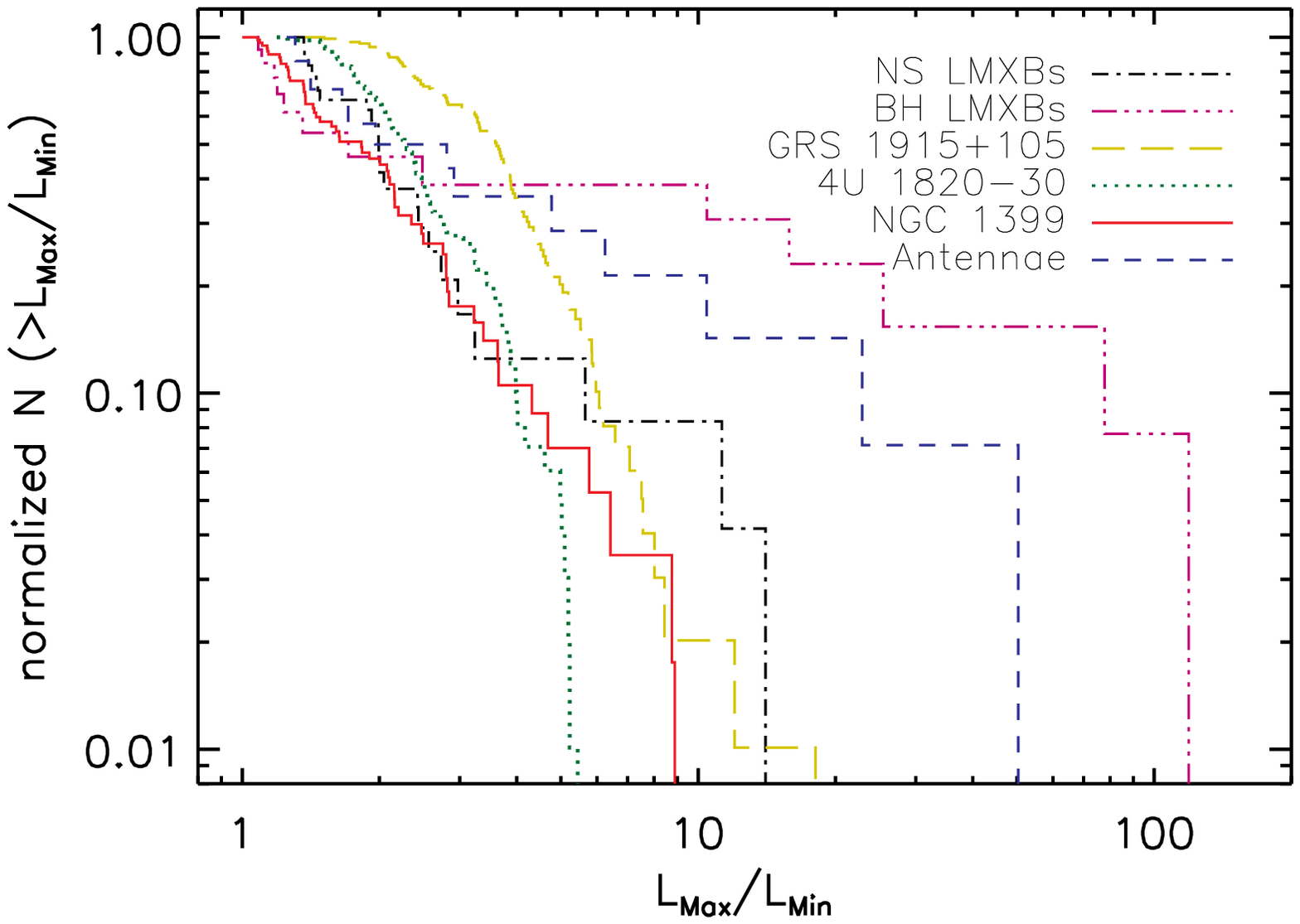} 
\caption{The X-ray
variability function of bright sources  ($L_{\rm max}>3\times10^{38}$
ergs s$^{-1}$) in NGC 1399 and the Antennae (top). Their normalized
variability functions are shown in the bottom panel to compare with
variability functions of 14 black hole LMXBs, 25 neutron star LMXBs,
the black hole LMXB GRS 1915$+$105 and the neutron star LMXB, 4U 1820$-$30, in a Galactic GC.
\label{fig:vfun}} 
\end{figure}

For all the X-ray sources in NGC 1399 and the Antennae, we calculated
their variability defined as described above. We defined the
variability function for a population as the number of sources with a
maximum variability larger than a certain value.  The variability
functions  for all bright sources ($L_{\rm X}>3\times10^{38}$ ergs
s$^{-1}$ in 0.3-10 keV) are shown in the top panel of
Fig.~\ref{fig:vfun}.  The fraction of bright sources which are variable
(with a confidence level of 90\% above the systematic error of
$\sqrt{2}$) is 10/63 for NGC 1399 and 9/15 for the Antennae.   The
fraction of ULXs which are variable is 1/8 for NGC 1399 and 4/4 for the
Antennae.  For NGC 1399, the only variable ULX is PSX-21.  The color of
the optical counterpart  ($C-R=0.69$, Washington $C$ and Kron $R$,
\citet{dir03}) for this source is inconsistent with that of a globular
cluster, but the color may be affected by a defect in the $R$-band
image (T. Richtler private communication).  Of the other 9 variable
sources, only two have optical counterparts within $1\arcsec$. 
Therefore, the majority of the variable sources are not associated with
globular clusters.  This is in sharp contrast to the fact that 70\% of
all sources in NGC 1399 are associated with globular clusters
\citep{ang01}. 

The detection threshold in all observations is well below our ULX
luminosity threshold.  However, the detection thresholds in the
shortest observations of NGC 1399 is close to our luminosity cutoff of
$3\times10^{38}$ ergs s$^{-1}$ for bright sources.  We examined two
long observations spaced by similar periods, i.e., (b) and (g) for NGC
1399 and (i) and (vii) for the Antennae, to check if the difference in
variability between NGC 1399 and the Antennae still persists when only
observations with detection thresholds well below our luminosity cutoff
are used.  We  found that 7 of the 63 bright sources in NGC 1399 and 4
of the 15 bright sources in the Antennae are variable.  This is
consistent with the conclusions that bright X-ray sources in NGC 1399
are less variable than those in the Antennae.

To compare with the variability function of X-ray binaries with known
characteristics, we selected 14 Galactic black hole LMXBs\footnote{ 
GRO~J0422$+$32, GRO~J1655$-$40, GRS~1009$-$45,  GS~1124$-$684,
GS~2000$+$250, GS~2023$+$338, GX~339$-$4, V4641~Sgr, X0620$-$003,
X1543$-$475, X1705$-$250, XTE~J1118$+$480, XTE~J1550$-$564,
XTE~J1859$+$226} and 25 neutron star LMXBs\footnote{ Cir~X-1, Cyg~X-2,
GS~1826$-$238, GX~13$+$1, GX~17$+$2, GX~3$+$1, GX~340$+$0,  GX~349$+$2,
GX~354$-$0, GX~5$-$1, GX~9$+$1, GX~9$+$9, LMC~X-2,  Sco~X-1, Ser~X-1,
X0614$+$091, X1254$-$690, X1543$-$624, X1608$-$522, X1705$-$440,
X1735$-$444, X1812$-$121, X1820$-$303, X1822$-$000, X2127$+$119} and
calculated their variability functions using data from the All Sky
Monitor (ASM) onboard the {\it Rossi X-ray Timing Explorer} ({\it
RXTE}).  For each source, we got one variability value by sampling the 
ASM one day average lightcurve spaced in the same way as those seven
{\it Chandra} observations of NGC 1399. Then we shifted the first date
to 10 days later and got another value until there were 100 values. The
variability of the source was derived from an average of these 100
values and then the variability functions were obtained, respectively,
for sources of black hole and neutron star LMXBs. We sampled the ASM
data spaced in the same way as the Antennae observations and the
results looked similar.  We also examined the variability of two
individual Galactic sources: GRS~1915$+$105 and 4U 1820$-$30.  The
Galactic black hole binary GRS~1915$+$105 has a luminosity in the ULX
range and is unusual in that it has been a persistent X-ray emitter for
14 years \citep{don04}.  It is also the most massive stellar mass black
hole that has been found so far. The bright neutron star LMXB 4U
1820$-$30 ($=$X1820$-$303) residing in the Galactic GC NGC 6624 has a
similar environment with sources in NGC 1399 and is an ultracompact
binary, though its luminosity is low relative to the sources considered
here. The variability function for each of these sources was obtained
from 100 ASM flux measurements.  The normalized variability functions
of selected black hole and neutron star LMXBs, GRS~1915$+$105, 4U
1820$-$30, NGC 1399 and the Antennae are shown together in the bottom
panel of Fig.~\ref{fig:vfun}.

\begin{figure}[b]
\centering
\plotone{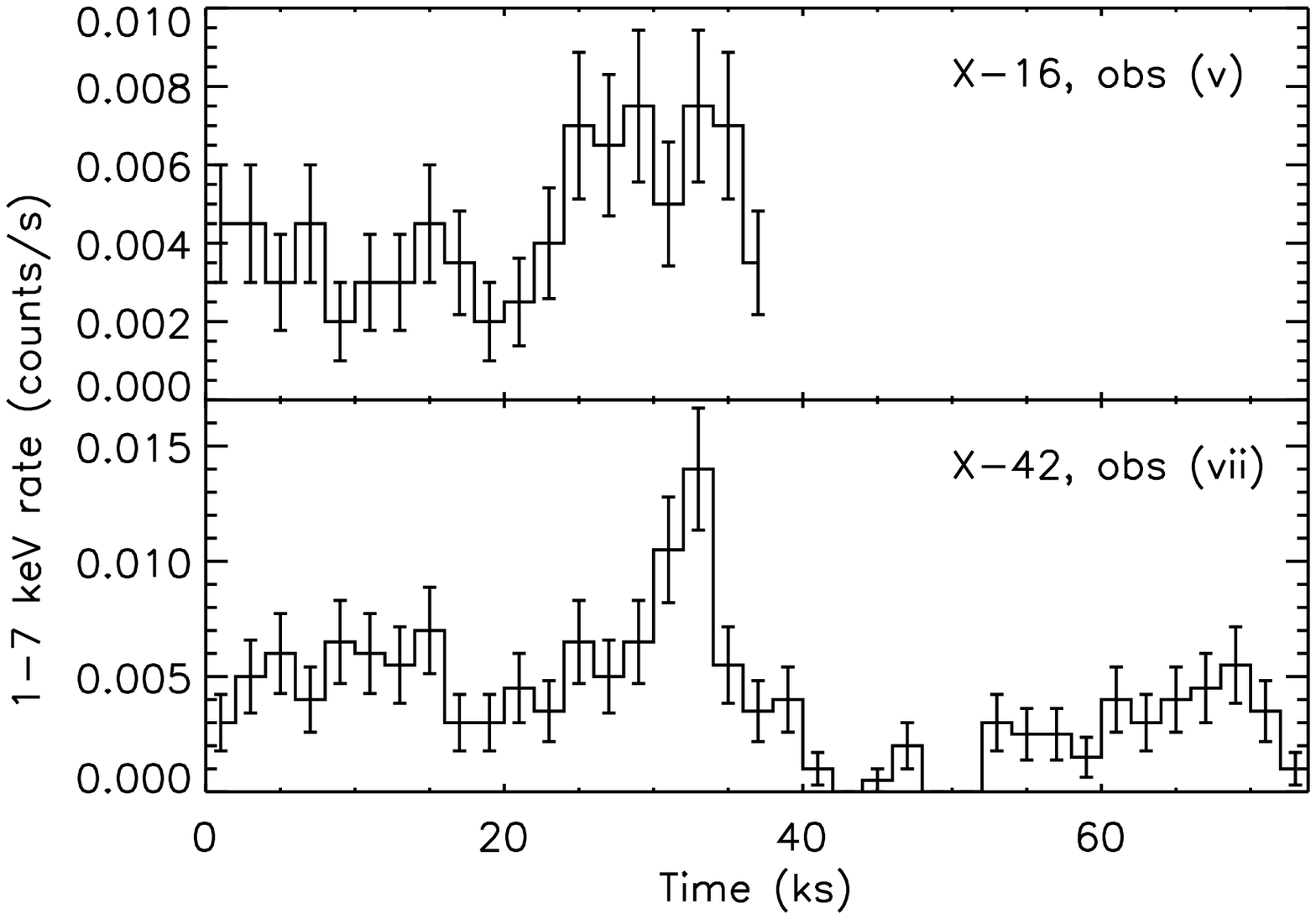}\\
\plotone{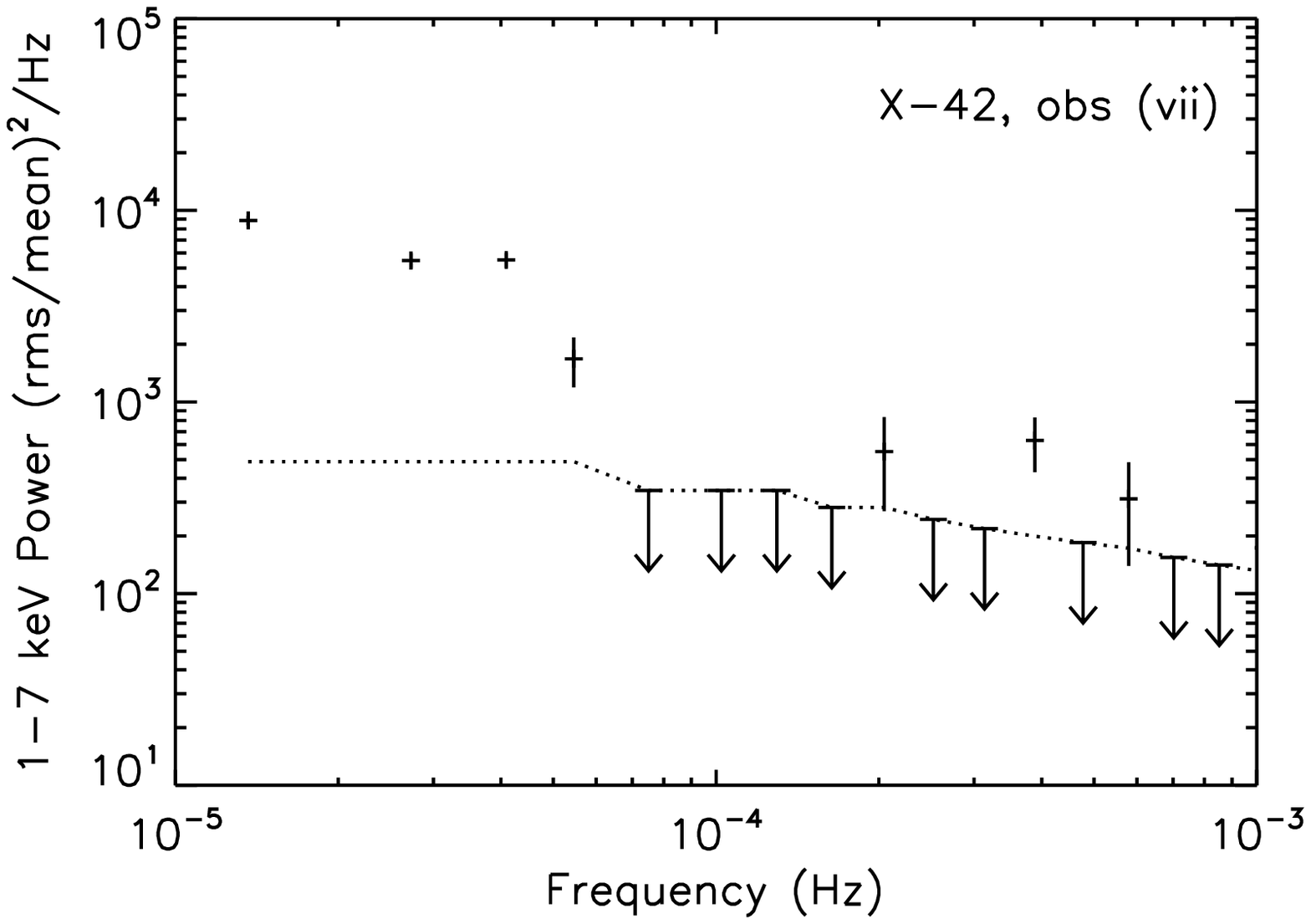}
\caption{Lightcurves for the Antennae X-16 in the observation (v) and X-42 in the
observation (vii) and the PSD for X-42. The top panel shows lightcurves in 1-7
keV band binned with 20 ks. The bottom panel presents the PSD in the 1-7 keV band
calculated from a 512-point FFT logarithmically binned by a factor of 1.2. The
dotted line indicates the Poisson level, and points below which are plotted with
2$\sigma$ upper limits. 
\label{fig:psd}}
\end{figure}

To estimate the typical outburst duration of sources in NGC 1399, we
used the two deepest observations of NGC 1399.  These are (b) and (g)
and are separated by $\Delta T=3.35$ yrs.  If a source was detected in
both observations, we assume the source lasted in its outburst state
for $\Delta T$.  We found 60 bright sources and 8 ULXs had at least one
detection in either (b) or (g) ,and among them 50 bright sources and 7
ULXs had two. Assuming a fixed outburst duration $T_{\rm B}$ for X-ray
sources in NGC 1399, we estimated that $T_{\rm B}=\Delta T/(1-\eta)$,
where $\eta$ was the fraction of twice detected sources among sources
with at least one detection. We thus derived $T_{\rm B}=20.1$ yrs for
bright sources and $T_{\rm B}=26.8$ yrs for ULXs.

To search for short-term variability for every X-ray source in each
observation, we used the K-S test to examine if the 0.3-10 keV X-ray
count rate is constant.  No source in NGC 1399 shows any short-term
variability above a significance level of 3$\sigma$. However, four
Antennae X-ray sources, including two ULXs, are variable with a
significance level above 3$\sigma$.  A 3.8$\sigma$ probability of
non-constant emission is shown for the Antennae X-16 in observation
(v), and an 8.0$\sigma$ level for X-42 in observation (vii).  The total
number of trials, equal to the number of detected sources multiplied by
the number of observations is 105.  Taking into account the number of
trials, the variability from X-16 is significant at the 98.7\% level,
while the variability from X-42 is highly significant with a chance
probability of occurrence of $4.1 \times 10^{-7}$. We examined the
light curves in several bands and found that the variability is most
pronounced in the 1-7~keV band; there is no significant variability
below 1~keV.  The 1-7 keV lightcurves for X-16 and X-42 and the power
spectrum density (PSD) for X-42 are shown in Fig.~\ref{fig:psd}. Strong
variability at a time scale around 20 ks in X-42 is explicit. The PSD
looks like a flat form that breaks at $4\times10^{-4}$ Hz, which is
similar to the break PSDs of Galactic black holes.  However, the power
spectrum is dominated by a single flare with a duration of about 5~ks
about halfway through the observation.  Therefore, it would be
imprudent to over interpret the power spectrum and attempt to derive a
mass for the compact object.  We note that no significant power is
detected in any other observation.

\subsection{X-ray Spectra for ULXs} \label{sec:spec}

\begin{figure}[h]
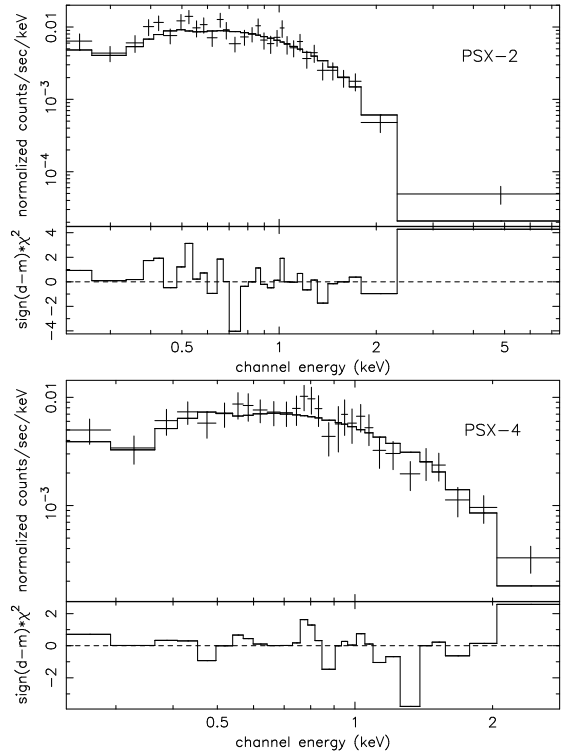

\centering
\plotone{f7a.eps}\\
\plotone{f7b.eps}
\caption{X-ray spectra of the source PSX-2 (top) and PSX-4 (bottom) in NGC 1399 from
observation (b) fitted by a {\tt diskbb} spectrum. The referred disk temperatures
are 0.40 keV and 0.42 keV, respectively for PSX-2 and PSX-4. However, both data sets
can also be fitted by a power-law model with a higher absorption. See
Table~\ref{tab:psx} for all fitting parameters. 
\label{fig:diskbb}}
\end{figure}

We examined all the X-ray spectra for ULXs in both galaxies with XSPEC
11.3.1. All sources in NGC 1399 and the Antennae could be reasonably
described by an absorbed power-law, whereas some sources in the
Antennae may show line features but with low significance (see
\cite{zez02-2}). Two sources (PSX-2 and PSX-4) in NGC 1399 present an
alternative favored fitting with a multicolor disk (MCD) blackbody
({\tt diskbb} in XSPEC) model (see Fig.~\ref{fig:diskbb}). All fittings
adopt an absorption with solar abundance ({\tt wabs} in XSPEC) as a
free parameter but setting the lower bound to the Galactic value. 

\begin{figure}[!h]
\centering
\plotone{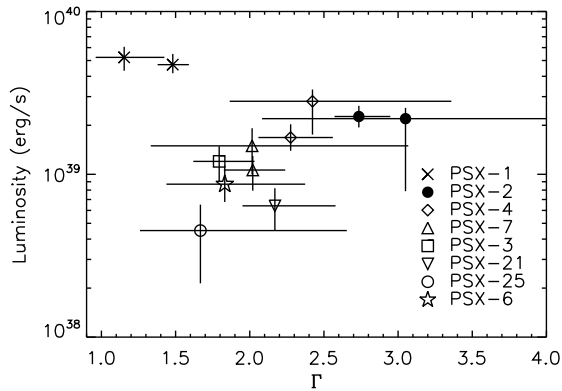}
\caption{0.3-10 keV luminosity vs. power-law photon index ($\Gamma$) for the 8 ULXs
in NGC 1399. All points in the plot come from observations that have enough photons
for a spectral fitting. PSX-1, PSX-2, PSX-4 and PSX-7 have two points from
observations (b) and (g), and rest sources have one from the observation (b). The
absorption in the fitting is set as a free parameter but with the lower bound to the
Galactic value ($1.34\times10^{20}$ cm$^{-2}$). The best fitted $n_{\rm H}$ are all
close or equal to the Galactic value (less than 3$\sigma$ except PSX-2 in observation (b), which is nearly
6$\sigma$). 
\label{fig:nlg}}
\end{figure}

Fitting parameters of sources PSX-2 and PSX-4 are shown in
Table~\ref{tab:psx}.  The best fitted 0.3-10 keV luminosity versus
photon index in every observation that contains enough photons are
plotted in Fig.~\ref{fig:nlg} for NGC 1399 ULXs and Fig.~\ref{fig:alg}
for the Antennae ULXs.  The best fitted column densities for ULXs in
NGC 1399 are consistent within 3$\sigma$ with the Galactic value,
except for PSX-2 in observation (b) which differs by nearly 6$\sigma$. 
The best fitted column densities for ULXs in the Antennae are shown on
each plot panel in Fig.~\ref{fig:alg}.  All photon indices and
luminosities in Fig.~\ref{fig:nlg} \& \ref{fig:alg} are obtained by
fitting a single power-law model with absorption.

\begin{deluxetable}{lllll}[!h]
\tablecaption{Spectral parameters with 90\% errors of PSX-2 and PSX-4 in NGC 1399 \label{tab:psx}}
\tablehead{
\colhead{source} & \colhead{model} & \colhead{$n_{\rm H}$ (cm$^{-2}$)} & \colhead{$kT$ (keV) or $\Gamma$} & \colhead{$\chi^2/{\rm dof}$}}

\startdata
PSX-2 & diskbb & $1.3_{-0.0}^{+0.7}\times10^{20}$ & $0.40_{-0.05}^{+0.02}$ & $30.6/29$\\
& powerlaw & $6.1_{-1.2}^{+1.4}\times10^{20}$ & $2.7_{-0.1}^{+0.2}$ & $37.7/29$\\

\noalign{\smallskip}\hline\noalign{\smallskip}

PSX-4 & diskbb & $1.3_{-0.0}^{+1.1}\times10^{20}$ & $0.42_{-0.07}^{+0.03}$ & $18.7/26$\\
& powerlaw & $4.2_{-1.5}^{+1.6}\times10^{20}$ & $2.3_{-0.2}^{+0.3}$ & $20.2/26$\\
\enddata
\end{deluxetable}

In Fig.~\ref{fig:nlg}, a weak correlation between the luminosity and
the power-law photon index is found for all ULXs except PSX-1 in NGC
1399 with a correlation coefficient of 0.66 and a probability of 3.7\%
for a coincidence from uncorrelated data. This correlation does not
exist in all ULXs in the Antennae, but in observations for individual
sources X-11, X-44 and X-42, the correlation coefficients with chance
probabilities in brackets are 0.61(14.6\%), $-$0.57(17.5\%) and
$-$0.66(10.6\%), respectively. These correlations are weak and more
observations are needed to expand the samples.

\section{Discussion} \label{sec:diss}

The most dramatic difference between NGC 1399 and the Antennae is that
the ULXs as well as the bright X-ray sources in the elliptical galaxy
are less variable.  In Fig.~\ref{fig:lln}, only a couple sources show
apparent variability in the L-L diagram of NGC 1399 between (b)-(e) or
(b)-(g) observations.  In Fig.~\ref{fig:lla}, the number of sources in
the Antennae with significant variability is obviously larger than that
in NGC 1399.  This is verified by lightcurves of ULXs and the
variability functions of bright sources.  

The variability functions of Galactic LMXBs show that most black
holes are highly variable, with the notable exception of GRS
1915$+$105, whereas bright neutron stars are less variable. The
normalized variability function (bottom panel of Fig.~\ref{fig:vfun})
presents the bright sources in NGC 1399 as similar to 4U 1820$-$30, a
little less variable than the group of neutron star LMXBs and GRS
1915$+$105, and far less  variable than the group of black hole
LMXBs.  ULXs in NGC 1399 are even less variable than other bright
sources, with 7 of 8 showing constant flux and the other having little
variability.  None of the bright X-ray sources or ULXs in NGC 1399
show short-term variability with a significance level of 3$\sigma$ in
any individual observation.  The lack of variability should provide a
strong constraint on the nature of these sources.

\begin{figure}[!h]
\centering
\plotone{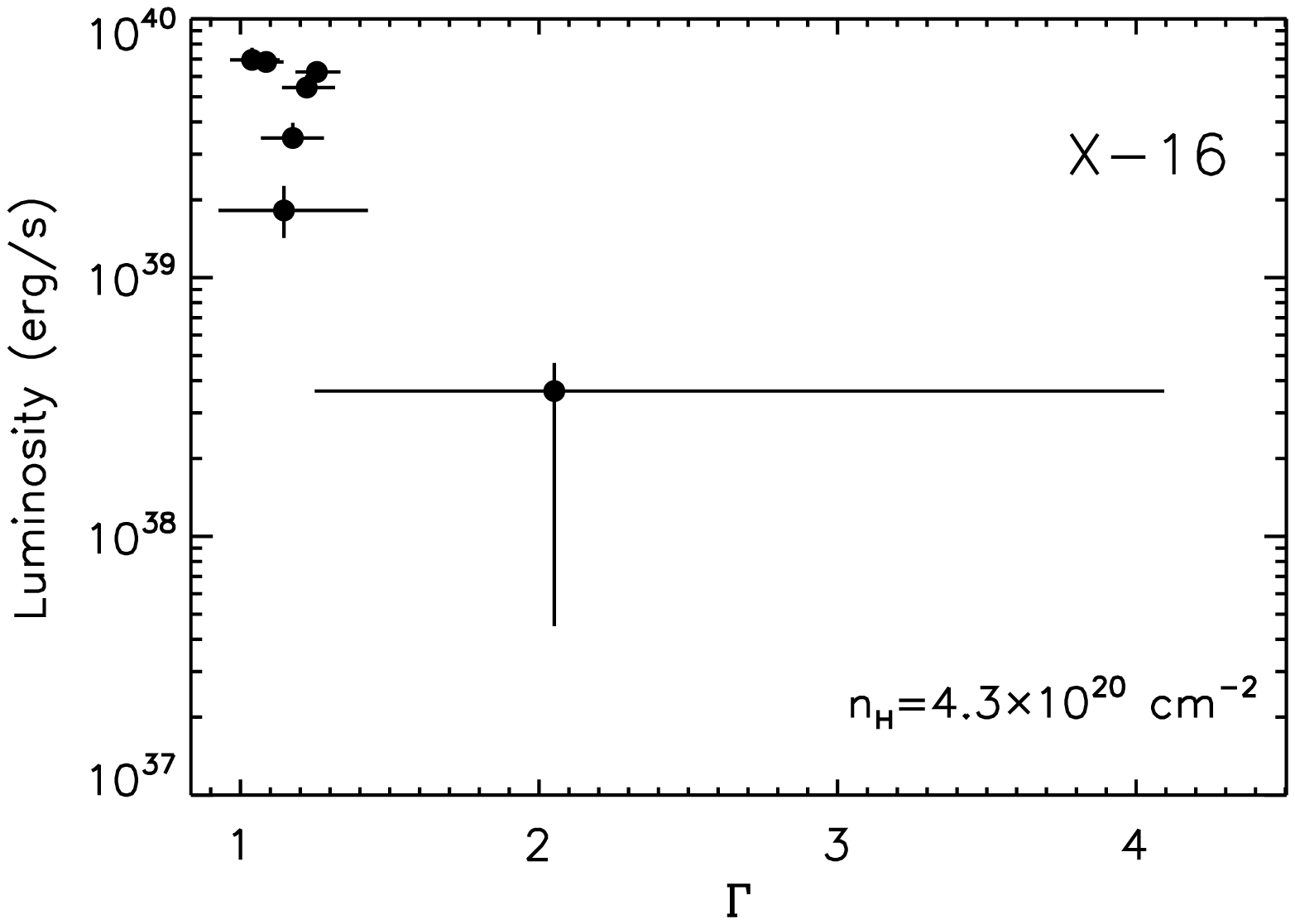}\\
\plotone{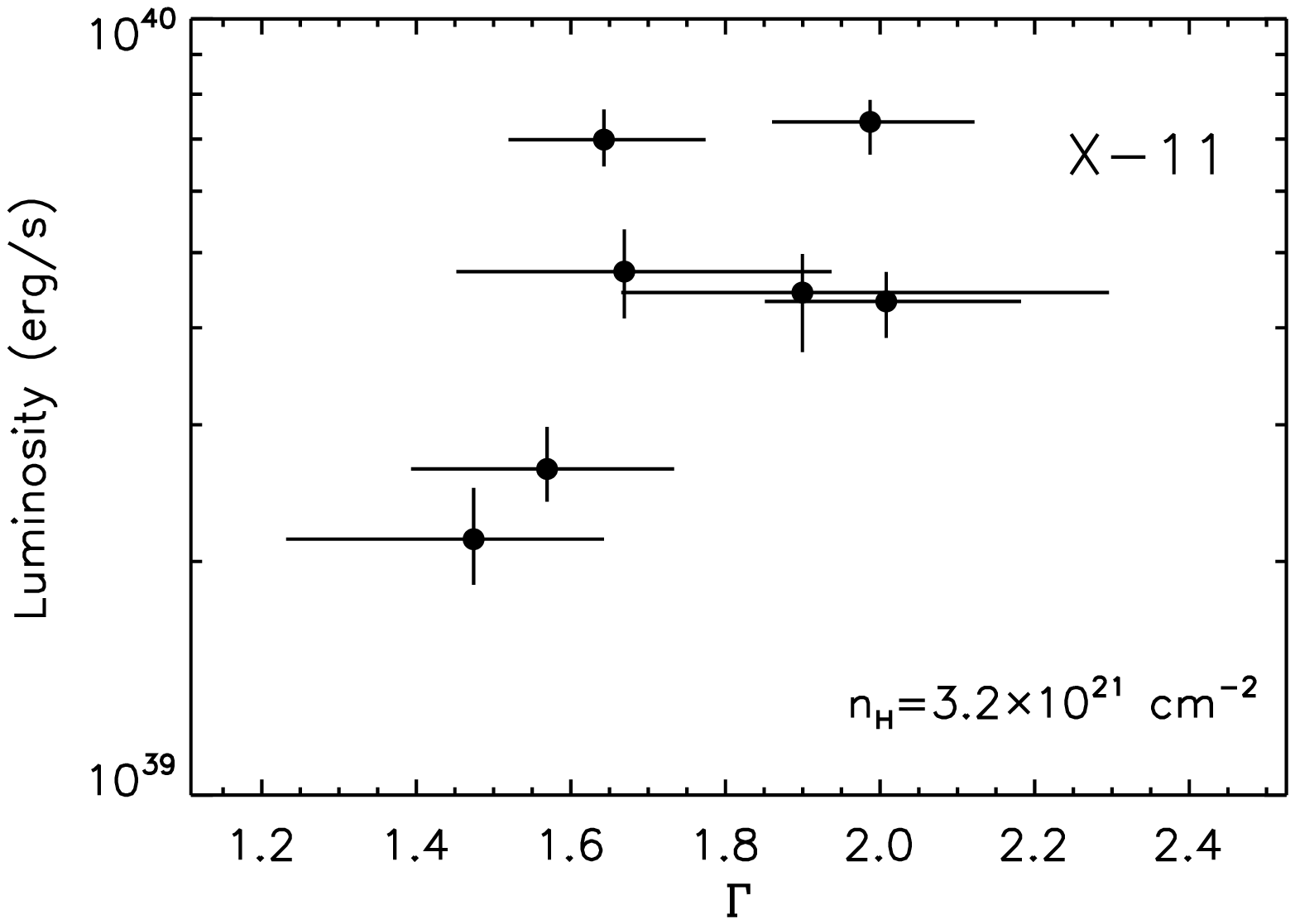}\\
\plotone{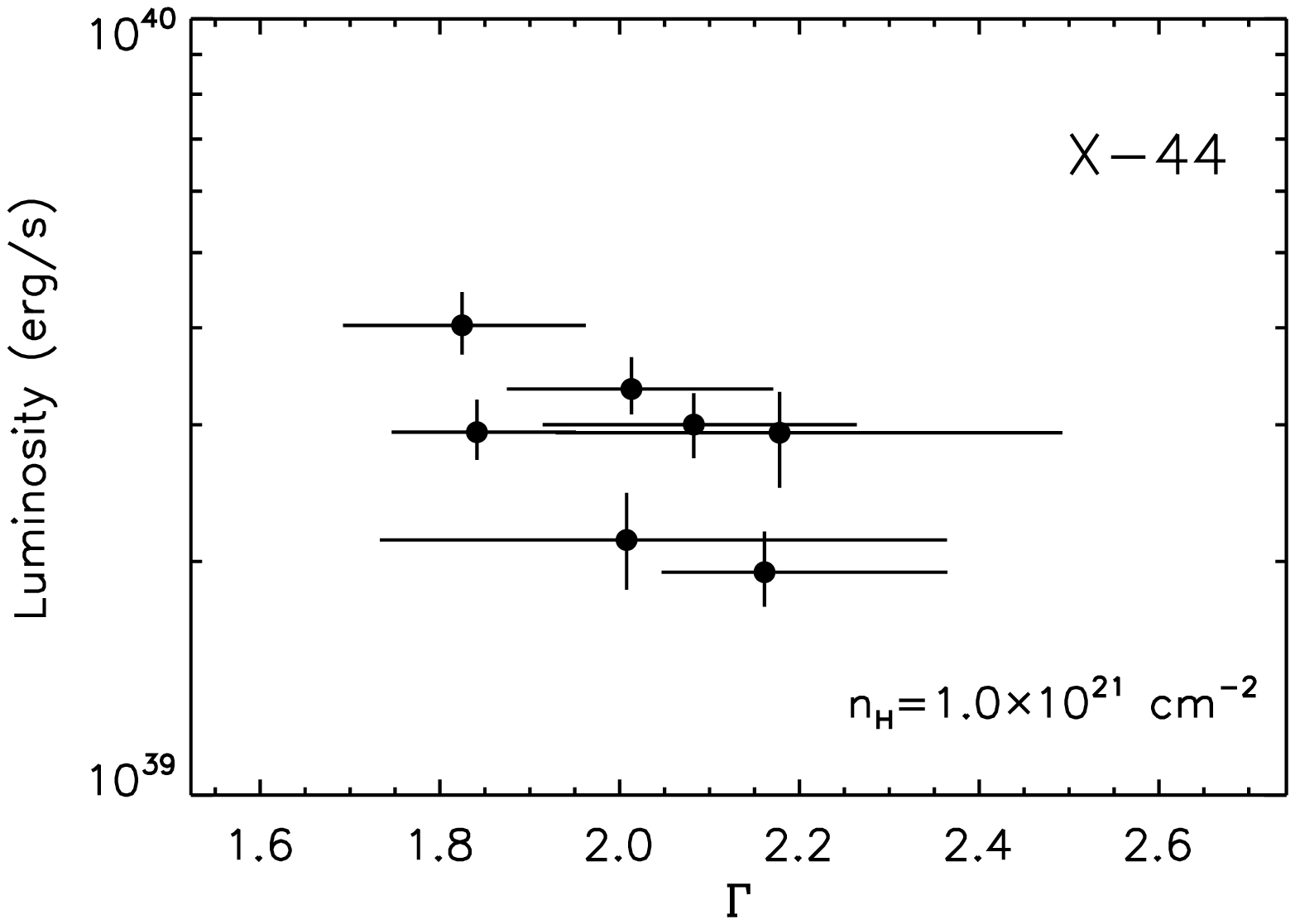}\\
\plotone{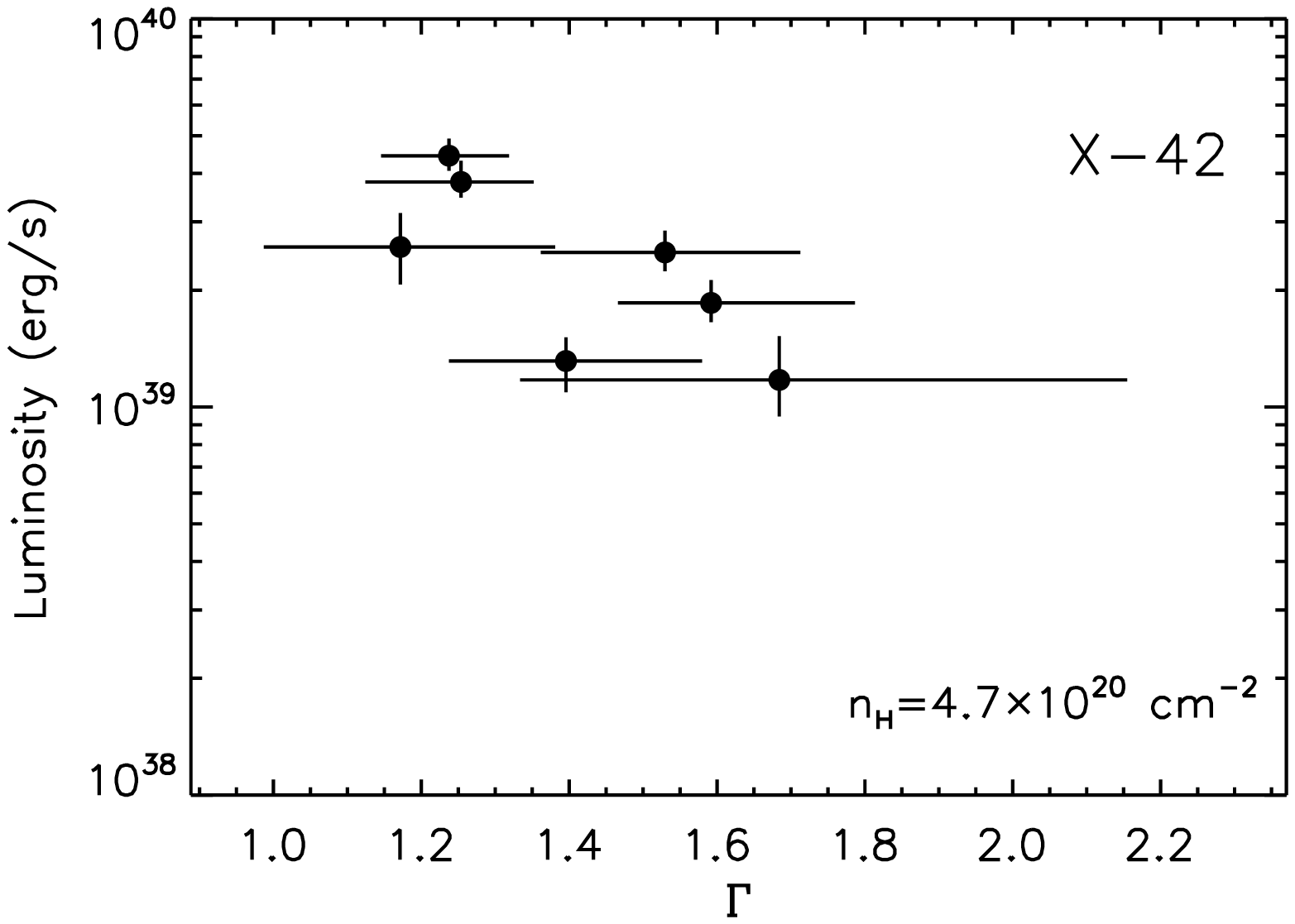}
\caption{0.3-10 keV luminosity vs. power-law photon index ($\Gamma$) for the 4 ULXs
in the Antennae in seven observations. The absorption in the fitting is set as a
free parameter but with the lower bound to the Galactic value ($3.83\times10^{20}$
cm$^{-2}$). The best fitted $n_{\rm H}$ values are consistent in all observations
and the weighted average number is shown for each source.  
\label{fig:alg}}
\end{figure}

The long term variability also allow us to estimate the outburst
duration of X-ray sources in NGC 1399.  We found that the outburst
duration is around 20 yrs for both bright X-ray sources and ULXs.

A natural way to produce low variability would be if the bright X-ray 
sources in NGC 1399 consist of multiple unresolved LMXBs.  However, a
large number of sources is required.  From the ASM data, we estimated
that neutron star LMXBs are  above 10\% of their maximum flux 80\% of
the time, while black holes are above their 10\% of their maximum flux
only 15\% of time (the estimate for black holes is an upper bound
because several of our selected black hole systems have had only one
outburst during the lifetime of RXTE).  Using the average luminosity of
4U 1820$-$30 of $\sim 5\times 10^{37} \rm \, erg \, s^{-1}$, about 8
neutron star binaries would be needed to produce the least luminous of
our bright sources and 50 would be needed to produce the least luminous
ULX.  Large numbers of neutron stars are known to be present in some
globular clusters in our galaxy, e.g.\ there are around 50 in the
Galactic GC NGC 6266 \citep{poo03}, but only very rarely does a cluster
contain more than one bright X-ray source.  It is very puzzling to
understand how individual clusters in NGC 1399 could contain so many
bright X-ray sources.  Though the luminosity of black holes in the 
accreting phase is, generally, several times larger than that of
neutron stars, the smaller duty cycle partially offsets the increased
luminosity and comparable numbers of black hole binaries are needed to
reach the same luminosity.  

Perhaps, the most promising interpretation of the NGC 1399 bright
sources is that they are are individual black holes of a type similar
to GRS 1915$+$105.  As noted above, GRS 1915$+$105 has been a
persistent X-ray source since its discovery 14 years ago \citep{cas94}
and has reached luminosities in the ULX range.  The variability
function of GRS 1915$+$105 is similar to that of the NGC 1399 bright
sources.  The companion star in GRS 1915$+$105 is a K or M giant
\citep{gre01}.  The time scale for the evolution of such a binary
system to the mass-transferring phase is roughly 6~Gyr \citep{bel02}. 
Allowing for different companion star masses, similar mass transferring
binaries could currently be present in either the old ($\sim 11$~Gyr)
or young ($\sim 2$~Gyr) globular clusters in NGC 1399 \citep{for01}. 
\citet{kin02} has previously suggested that the ULXs in ellipticals may
be similar to GRS 1915$+$105.  \citet{pir02} suggested that the bright
X-ray sources in ellipticals are X-ray binaries with Roche-lobe filling
giant companions with mass transfer driven by nuclear evolution of the
companion.  They predicted outburst durations similar to what we find
for NGC 1399.  For the sources associated with globular clusters,
another possibility is that the companion stars are initially isolated
stars which are captured by the black hole after evolving to the giant
phase.  The capture cross-section is enhanced for stars in the giant
phase.

However, GRS 1915$+$105 shows strong variability on short time
scales, which we do not observe from the bright X-ray sources in NGC
1399.  The sensitivity of Chandra to short timescale variability is
limited due to the low counting rate, but our detection of short
timescale variability from sources in the Antennae (with comparable
counting rates) indicates that similar flares would have been detected
from the NGC 1399 sources.  The short timescale variability of sources
in NGC 1399 should be investigated in more detail with new
observations.

\citet{bil04} has suggested that the bright X-ray sources in NGC
1399 may be neutron stars accreting from white dwarf companions in
ultracompact binaries, similar to 4U 1820$-$30.  The variability of 4U
1820$-$30 is similar to the NGC 1399 bright sources. Hence, neutron
star ultracompact binaries may account for some or all of the bright
sources if mildly super-Eddington (apparent) luminosities can be
achieved.  However, the short lifetimes expected for the high accretion
rates required given the very low-mass companions ($0.04 - 0.08
M_{\odot}$) make it unlikely that the ULXs in NGC 1399 are neutron star
ultracompact binaries.  Another possible interpretation of the bright
sources is that they are neutron-star binaries similar to the
neutron-star X-ray binaries known as Z sources.  Z sources are
persistent with variability similar to that seen from the NGC 1399
sources.  Again, mildly super-Eddington (apparent)
luminosities would be required to produce the dimmest of the bright
sources and such objects are unlikely to explain the ULXs. 
Interestingly, \citet{web83} suggested that the Z sources have giant
star companions with mass loss driven by nuclear evolution.  This would
make the systems the neutron star analogs of black hole/giant star
binaries such as GRS 1915$+$105.

The three most luminous X-ray sources in NGC 1399 are interesting. The
brightest one, PSX-1, has a hard spectrum ($\Gamma=1.2\sim1.5$), little
variability ($L_{\rm max}/L_{\rm min}=2.2\pm0.8$), and a high
luminosity around $5\times10^{39}$ ergs s$^{-1}$.  The spectrum
indicates the source is not in its high/soft state ($\Gamma=2.1-4.8$)
but at the hard extreme of the low/hard state ($\Gamma=1.5-2.1$)
\citep{mcc06}.  The low variability indicates that the source is
probably not relativistically beamed \citep{kor02}, because
relativistically beamed sources are expected to be highly variable.
This source is associated with a bright GC with $C-R=2.12$, $m_R=21.02$
\citep{dir03}, which means it is unlikely to be a background AGN.  A
huge number of sources would be required to explain the emission as a
collection of neutron star or black hole X-ray binaries.  If the source
is an IMBH in the low state, its practically constant luminosity is
reasonable if compared to the long-term behavior of M82 X-1, which
varies in luminosity by a factor of 2-3 \citep{kaa06a,kaa06b} and shows
a hard spectrum with a power-law index $\Gamma \approx 1.6$
\citep{kaa06b}.

NGC 1399 PSX-2 and PSX-4 display typical behaviors of black holes in
the high/soft state.  PSX-2 is associated with a bright GC  ($C-R =
2.24$, $m_R = 21.36$) but PSX-4 is not. The plausible cool disk ($kT
\approx 0.4$ keV) emission is consistent with the soft spectral index
($\Gamma \approx 2.5$).  Their low variability ($L_{\rm max}/L_{\rm
min} = 2.5 \pm 0.7$ for PSX-2 and $1.5 \pm 0.6$ for PSX-4) is
suggestive of some Galactic black holes in the high state, but no
Galactic source has been found to remain in the high state for four
years \citep{mcc06}.  The cool disk has been a useful probe in finding
IMBH candidates \citep{col99,kaa03,mil04,fen05}.  However, their disk
temperatures are somewhat higher than most IMBH candidates, and their
luminosities ($2\sim3\times10^{39}$ ergs s$^{-1}$) are also around the
lower bound of those candidates \citep{mil04,fen05}.  These sources do
not have high mass companion stars to produce geometrically beamed
emission as suggested by \citet{kin01}.  Therefore, it could be
possible that these two sources are 10$\sim$100$M_\sun$ black holes
emitting at the high state if the disk components are true.  One
possible way to determine the reality of the disk emission is an
accurate measurement of the column density, since these two sources, in
particular PSX-2, show significantly higher intrinsic absorption if
modeled by a single power-law spectrum.

The bright X-ray sources in the Antennae are variable.  This suggests
that they are black hole binaries and accreting from massive stars to
produce unstable disks \citep{kal04}.  They are not as variable as
black hole LMXBs (Fig.~\ref{fig:vfun}).  Considering the variability
together with their high luminosity, we suggest they are HMXBs
accreting via Roche-lobe overflow.  There is no disk emission found in
any of the 4 ULXs.

The most luminous source, X-16, manifests a very hard spectra
($\Gamma=1.0-1.3$) and has strong variability with a change of
luminosity of $\sim$10 (Fig.~\ref{fig:alg}).  The source could be an
IMBH with a mass of a few hundred to thousand $M_\sun$ emitting at the
hard state.  The X-ray behavior is also consistent with a
relativistically beamed jet pointing at us, but \citet{zez02-3} pointed
out that there is no radio emission found from any ULX in the
Antennae.  It is interesting that none of the ULXs shows the very soft
spectrum normally seen in ULXs in nearby galaxies \citep{fen05}.  We
note that X-37 was reported to have a soft component in its spectrum,
but has been subsequently identified as a background AGN.  X-11 and
X-42 are likely to be in the hard state ($\Gamma \le 2$), X-44 looks
like it is between the hard and soft states ($\Gamma \sim 2$).  A
plausible correlation ($r=0.61$) is found in X-11 between luminosity
and $\Gamma$, which shows similar behaviors as stellar mass black
holes, but we should be cautious because of the weak correlation
coefficient.  The X-ray properties of these three sources are
consistent with HMXBs, but whether they are IMBHs or $\sim$20$M_\sun$
black holes emitting in near or super Eddington luminosities is still
unclear.

\acknowledgments 

We thank Tom Richtler for providing us the GC list in NGC 1399 and
giving instructions on the GC identification and an anonymous referee
for comments which improved our interpretation of the results.  HF
thanks Zach Prieskorn for helping correct the English in the paper.  We
acknowledge support from NASA grants HST-GO-10001 and Chandra grant
GO4-5086A. PK acknowledges support from a University of Iowa Faculty
Scholar Award.

\end{document}